\documentclass[pra,preprint,showpacs,floatfix]{revtex4}
\usepackage{amsmath}
\usepackage{amsfonts}
\usepackage{amssymb}
\usepackage{dcolumn}
\usepackage{bm}
\usepackage{graphicx}
\newcommand{\graphr}{
\setlength{\unitlength}{1.8ex}
\begin{picture}(1.05,1)
\linethickness{1pt}
\put(.5,0){\circle*{.2}}
\put(0.5,.5){\circle{1}}
\end{picture}
}

\newcommand{\graphgamma}{
\setlength{\unitlength}{1.8ex}
\begin{picture}(1,1)(0,-0.25)
\linethickness{.5pt}
\put(0,0){\line(1,0){1}}
\put(0,0){\circle*{.2}}
\put(1,0){\circle*{.2}}
\end{picture}
}

\newcommand{\graphrra}{
\setlength{\unitlength}{1.8ex}
\begin{picture}(2,.5)(-.5,-0.25)
\put(0.5,0){\circle*{.2}}
\linethickness{.5pt}
\put(0,0){\circle{1}}
\put(1,0){\circle{1}}
\end{picture}
}

\newcommand{\graphrrb}{
\setlength{\unitlength}{1.8ex}
\begin{picture}(3,.5)(-0.25,.1)
\linethickness{1pt}
\put(0.50,0.00){\circle*{.2}}
\put(0.50,0.50){\circle{1}}

\put(2,0.00){\circle*{.2}}
\put(2,0.50){\circle{1}}

\end{picture}
}

\newcommand{\graphgra}{
\setlength{\unitlength}{1.8ex}
\begin{picture}(2.5,.5)(-0.25,-0.25)
\put(1,0){\circle*{.2}}
\put(2,0){\circle*{.2}}
\linethickness{1pt}
\put(0.5,0){\circle{1}}
\linethickness{.5pt}
\put(1,0){\line(1,0){1}}
\end{picture}
}

\newcommand{\graphgrb}{
\setlength{\unitlength}{1.8ex}
\begin{picture}(3.5,.5)(-0.25,-0.25)
\linethickness{1pt}
\put(1,0){\circle*{.2}}
\put(2,0){\circle*{.2}}
\put(3,0){\circle*{.2}}
\put(0.5,0){\circle{1}}

\linethickness{.5pt}
\put(2,0){\line(1,0){1}}
\end{picture}
}

\newcommand{\graphgga}{
\setlength{\unitlength}{1.8ex}
\begin{picture}(1.5,.5)(-0.25,-0.25)
\linethickness{.5pt}
\put(0,0){\circle*{.2}}
\put(1,0){\circle*{.2}}
\linethickness{1pt}
\put(0.5,0){\circle{1}}
\end{picture}
}

\newcommand{\graphggb}{
\setlength{\unitlength}{1.8ex}
\begin{picture}(1.5,.5)(-0.25,-0.25)
\linethickness{.5pt}
\put(0,-0.5){\circle*{.2}}
\put(0,-0.5){\line(1,0){1}}
\put(1,-.5){\circle*{.2}}
\put(1,-0.5){\line(0,1){1}}
\put(1,0.5){\circle*{.2}}
\end{picture}
}

\newcommand{\graphggc}{
\setlength{\unitlength}{1.8ex}
\begin{picture}(1.5,.5)(-0.25,-0.25)
\linethickness{.5pt}
\put(0,0.5){\line(1,0){1}}
\put(0,-0.5){\line(1,0){1}}
\linethickness{.5pt}
\put(0,0.5){\circle*{0.2}}
\put(0,-0.5){\circle*{0.2}}
\put(1,0.5){\circle*{0.2}}
\put(1,-0.5){\circle*{0.2}}
\end{picture}
}

\newcommand{\graphrrra}{
\setlength{\unitlength}{1.5ex}
\begin{picture}(3,1.5)(-1.5,-0.5)
\linethickness{0.001em}
\put(0,0){\circle*{0.5}}
\qbezier(0.00,0.00)(0.281,1.875)(1.125,0.750)
\qbezier(1.125,0.750)(1.969,-0.375)(0.00,0.00)
\qbezier(0.00,0.00)(-1.406,-1.125)(0.00,-1.125)
\qbezier(0.00,-1.125)(1.406,-1.125)(0.00,0.00)
\qbezier(0.00,0.00)(-1.969,-0.375)(-1.125,0.750)
\qbezier(-1.125,0.750)(-0.281,1.875)(0.00,0.00)
\end{picture}}

\newcommand{\graphrrrb}{
\setlength{\unitlength}{1.8ex}
\begin{picture}(3,1)(-0.25,-0.25)
\linethickness{1pt}
\put(0.5,0){\circle*{.2}}
\linethickness{1pt}
\put(2,0){\circle*{.2}}
\put(0.5,-0.5){\circle{1}}
\put(0.5,0.5){\circle{1}}
\put(2,0.5){\circle{1}}
\end{picture}
}

\newcommand{\graphrrrc}{
\setlength{\unitlength}{1.8ex}
\begin{picture}(4.50,1)(-0.25,0)
\linethickness{1pt}
\put(0.50,0.00){\circle*{.2}}
\put(0.50,0.50){\circle{1}}

\put(2,0.00){\circle*{.2}}
\put(2,0.50){\circle{1}}

\put(3.50,0.00){\circle*{.2}}
\put(3.50,0.50){\circle{1}}
\end{picture}
}

\newcommand{\graphgrra}{
\setlength{\unitlength}{1.8ex}
\begin{picture}(2,1)(-0.25,-0.25)
\linethickness{1pt}
\put(0.5,0){\circle*{.2}}
\put(1.5,0){\circle*{.2}}
\put(0.5,-0.5){\circle{1}}
\put(0.5,0.5){\circle{1}}

\linethickness{0.02em}
\put(0.5,0){\line(1,0){1}}
\end{picture}
}

\newcommand{\graphgrrb}{
\setlength{\unitlength}{1.8ex}
\begin{picture}(3.5,.75)(-0.25,-0.25)
\linethickness{1pt}
\put(1,0){\circle*{.2}}
\put(2,0){\circle*{.2}}
\put(0.5,0){\circle{1}}
\put(2.5,0){\circle{1}}

\linethickness{.5pt}
\put(1,0){\line(1,0){1}}
\end{picture}
}

\newcommand{\graphgrrc}{
\setlength{\unitlength}{1.8ex}
\begin{picture}(4,1)(-0.25,-0.25)
\linethickness{1pt}
\put(1,0){\circle*{.2}}
\put(2,0){\circle*{.2}}
\put(3,0){\circle*{.2}}
\put(0.5,0){\circle{1}}
\put(3,0.5){\circle{1}}

\linethickness{.5pt}
\put(1,0){\line(1,0){1}}
\end{picture}
}

\newcommand{\graphgrrd}{
\setlength{\unitlength}{1.8ex}
\begin{picture}(2.5,1)(-0.25,-0.25)
\linethickness{1pt}
\put(1,0.5){\circle*{.2}}
\put(0.5,-0.5){\circle*{.2}}
\put(1.5,-0.5){\circle*{.2}}
\put(0.5,0.5){\circle{1}}
\put(1.5,0.5){\circle{1}}

\linethickness{.5pt}
\put(0.5,-0.5){\line(1,0){1}}
\end{picture}
}

\newcommand{\graphgrre}{
\setlength{\unitlength}{1.8ex}
\begin{picture}(3,1)(-0.25,-0.25)
\linethickness{1pt}
\put(0.5,0){\circle*{.2}}
\put(2,0){\circle*{.2}}
\put(0.75,-0.5){\circle*{.2}}
\put(1.75,-0.5){\circle*{.2}}
\put(0.5,0.5){\circle{1}}
\put(2,0.5){\circle{1}}

\linethickness{.5pt}
\put(0.75,-0.5){\line(1,0){1}}
\end{picture}
}

\newcommand{\graphggra}{
\setlength{\unitlength}{1.8ex}
\begin{picture}(2.5,.75)(-0.25,-0.25)
\linethickness{1pt}
\put(0,0){\circle*{.2}}
\put(1,0){\circle*{.2}}
\put(0.5,0){\circle{1}}
\put(1.5,0){\circle{1}}
\end{picture}
}

\newcommand{\graphggrb}{
\setlength{\unitlength}{1.8ex}
\begin{picture}(3,.75)(-0.25,-0.25)
\linethickness{1pt}
\put(0,0){\circle*{.2}}
\put(1,0){\circle*{.2}}
\put(2,-0.5){\circle*{.2}}
\put(0.5,0){\circle{1}}
\put(2,0){\circle{1}}
\end{picture}
}

\newcommand{\graphggrc}{
\setlength{\unitlength}{1.8ex}
\begin{picture}(2.5,.75)(-0.25,-0.25)
\linethickness{.5pt}
\put(1,-0.25){\line(1,0){1}}
\put(1,-0.25){\line(-1,0){1}}

\put(1,-0.25){\circle*{.2}}
\put(0,-0.25){\circle*{.2}}
\put(2,-0.25){\circle*{.2}}
\linethickness{1pt}
\put(1,0.25){\circle{1}}
\end{picture}
}

\newcommand{\graphggrd}{
\setlength{\unitlength}{1.8ex}
\begin{picture}(2.5,.75)(-0.25,-0.25)
\linethickness{.5pt}
\put(0,0.75){\circle*{.2}}
\put(0,0.75){\line(0,-1){1}}
\put(0,-0.25){\circle*{.2}}
\put(0,-0.25){\line(1,0){1}}
\put(1,-0.25){\circle*{.2}}
\linethickness{1pt}
\put(1.5,-0.25){\circle{1}}
\end{picture}
}

\newcommand{\graphggre}{
\setlength{\unitlength}{1.8ex}
\begin{picture}(3,.75)(-0.25,-0.25)
\linethickness{.5pt}
\put(0,0.5){\circle*{.2}}
\put(0,0.5){\line(0,-1){1}}
\put(0,-0.5){\circle*{.2}}
\put(0,-0.5){\line(1,0){1}}
\put(1,-0.5){\circle*{.2}}
\put(2,-0.5){\circle*{.2}}
\linethickness{1pt}
\put(2,0){\circle{1}}
\end{picture}
}

\newcommand{\graphggrf}{
\setlength{\unitlength}{1.8ex}
\begin{picture}(2.5,1.25)(-0.25,-0.25)
\linethickness{.5pt}
\put(0,0.5){\line(1,0){1}}
\put(0,-0.5){\line(1,0){1}}
\linethickness{.5pt}
\put(0,0.5){\circle*{0.2}}
\put(0,-0.5){\circle*{0.2}}
\put(1,0.5){\circle*{0.2}}
\put(1,-0.5){\circle*{0.2}}
\put(1.5,0.5){\circle{1}}
\end{picture}
}

\newcommand{\graphggrg}{
\setlength{\unitlength}{1.8ex}
\begin{picture}(3,.75)(-0.25,-0.25)
\linethickness{.5pt}
\put(0,0.5){\line(1,0){1}}
\put(0,-0.5){\line(1,0){1}}
\linethickness{.5pt}
\put(0,0.5){\circle*{0.2}}
\put(0,-0.5){\circle*{0.2}}
\put(1,0.5){\circle*{0.2}}
\put(1,-0.5){\circle*{0.2}}
\put(1.5,0){\circle*{0.2}}
\put(2,0){\circle{1}}
\end{picture}
}

\newcommand{\graphggga}{
\setlength{\unitlength}{1.8ex}
\begin{picture}(1.5,1)(-0.25,-0.25)
\linethickness{.5pt}
\put(0,0){\circle*{.2}}
\put(0,0){\line(1,0){1}}
\put(1,0){\circle*{.2}}
\linethickness{1pt}
\put(0.5,0){\circle{1}}
\end{picture}
}

\newcommand{\graphgggb}{
\setlength{\unitlength}{1em}
\begin{picture}(1.5,.75)(-0.25,-0.25)
\linethickness{.5pt}
\put(0,-0.5){\circle*{.2}}
\put(0,-0.5){\line(1,1){1.0142}}
\put(0,-0.5){\line(1,0){1}}
\put(1,-.5){\circle*{.2}}
\put(1,-0.5){\line(0,1){1}}
\put(1,0.5){\circle*{.2}}
\end{picture}
}

\newcommand{\graphgggc}{
\setlength{\unitlength}{1.8ex}
\begin{picture}(2.5,.5)(-0.25,-0.25)
\linethickness{.5pt}
\put(0,0){\circle*{.2}}
\put(1,0){\circle*{.2}}
\put(1,0){\line(1,0){1}}
\put(2,0){\circle*{.2}}
\linethickness{1pt}
\put(0.5,0){\circle{1}}
\end{picture}
}

\newcommand{\graphgggd}{
\setlength{\unitlength}{1.8ex}
\begin{picture}(2.5,.75)(-0.25,-0.25)
\linethickness{.5pt}
\put(0,-0.5){\circle*{.2}}
\put(0,-0.5){\line(1,0){1}}
\put(1,-0.5){\circle*{.2}}
\put(1,-0.5){\line(1,0){1}}
\put(1,-0.5){\line(0,1){1}}
\put(1,0.5){\circle*{.2}}
\put(2,-0.5){\circle*{.2}}
\end{picture}
}

\newcommand{\graphggge}{
\setlength{\unitlength}{1.8ex}
\begin{picture}(1.5,1)(-0.25,-0.25)
\linethickness{.5pt}
\put(0,0.5){\circle*{.2}}
\put(0,0.5){\line(0,-1){1}}
\put(0,-0.5){\circle*{.2}}
\put(0,-0.5){\line(1,0){1}}
\put(1,-0.5){\circle*{.2}}
\put(1,-0.5){\line(0,1){1}}
\put(1,0.5){\circle*{.2}}
\end{picture}
}

\newcommand{\graphgggf}{
\setlength{\unitlength}{1.8ex}
\begin{picture}(1.5,1)(-0.25,-0.25)
\linethickness{.5pt}
\put(0,0.5){\circle*{.2}}
\put(0,-0.5){\circle*{.2}}
\put(1,0.5){\circle*{.2}}
\put(0,-0.5){\line(1,0){1}}
\put(1,-0.5){\circle*{.2}}
\linethickness{1pt}
\put(0.5,0.5){\circle{1}}\end{picture}
}

\newcommand{\graphgggg}{
\setlength{\unitlength}{1.8ex}
\begin{picture}(1.5,1)(-0.25,-0.25)
\linethickness{.5pt}
\put(0,0.75){\circle*{.2}}
\put(0,0.75){\line(1,0){1}}
\put(1,0.75){\circle*{.2}}
\put(1,0.75){\line(0,-1){1}}
\put(1,-0.25){\circle*{.2}}

\put(0,-0.75){\circle*{.2}}
\put(0,-0.75){\line(1,0){1}}
\put(1,-0.75){\circle*{.2}}
\end{picture}
}

\newcommand{\graphgggh}{
\setlength{\unitlength}{1.8ex}
\begin{picture}(1.5,1)(-0.25,-0.25)
\linethickness{.5pt}
\put(0,0.5){\circle*{.2}}
\put(0,0.5){\line(1,0){1}}
\put(1,0.5){\circle*{.2}}

\put(0,0){\circle*{.2}}
\put(0,0){\line(1,0){1}}
\put(1,0){\circle*{.2}}

\put(0,-0.5){\circle*{.2}}
\put(0,-0.5){\line(1,0){1}}
\put(1,-0.5){\circle*{.2}}
\end{picture}
}

\newcommand{\labeledgraphr}[1]{
\setlength{\unitlength}{1.8ex}
\begin{picture}(2,.75)(0,0.25)
\linethickness{1pt}
\put(1,.5){\circle*{.2}}
\put(1.5,0.25){\scriptsize{$#1$}}
\linethickness{1pt}
\put(0.5,0.5){\circle{1}}
\end{picture}
}

\newcommand{\labeledgraphgamma}[2]{
\setlength{\unitlength}{1.8ex}
\begin{picture}(2,.75)(-0.2,0)
\linethickness{1pt}
\put(0.25,0){\circle*{.2}}
\put(0.25,0){${}^{#1}$}
\put(1.25,0){\circle*{.2}}
\put(1.25,0){${}^{#2}$}
\linethickness{.5pt}
\put(0.25,0){\line(1,0){1}}
\end{picture}
}

\newcommand{\labeledgraphgra}[2]{
\setlength{\unitlength}{1.8ex}
\begin{picture}(2.5,1)(0,0)
\linethickness{1pt}
\put(0.5,1){\circle*{.2}}
\put(-0.5,1.5){\scriptsize $#1$}
\put(1.5,1){\circle*{.2}}
\put(2,0.25){\scriptsize $#2$}
\put(0.5,0.5){\circle{1}}
\linethickness{.5pt}
\put(0.5,1){\line(1,0){1}}
\end{picture}
}

\newtheorem{definition}{Definition}

\newcommand{\stackleft}[2]{ \,{}^{(#1)}_{\ #2}}
\newcommand{\stacklr }[5]{ \,{}^{(#1)}_{\ #2}#3^{#4}_{#5}\,}

\newcommand{\Rmn}[1]{\uppercase\expandafter{\romannumeral #1}}
\newcommand{\chir}{\protect\raisebox{0.4ex}{$\chi$}}
\newcommand{\bchir}{\protect\raisebox{0.4ex}{$\bm{\chi}$}}
\allowdisplaybreaks
\makeindex
\begin{document}
\title{A Test of a New Interacting $\bm{N}$-Body Wave Function}
\author{Martin Dunn, W. Blake Laing\footnote{Current Address: Department of Physics, Kansas State University},
Derrick Toth\footnote{Current Address: Department of Physics and Astronomy,
University of Minnesota}, and Deborah K. Watson}
\affiliation{Homer L. Dodge Department of Physics and Astronomy, University of Oklahoma}
\date{\today}

\begin{abstract}
The resources required to solve the general interacting quantum $N$-body problem scale exponentially with $N$, making the solution of this problem very difficult when $N$ is large.  In a previous series of papers we develop an approach for a fully-interacting wave function with a general two-body interaction which tames the $N$-scaling by developing a perturbation series that is order-by-order invariant under a point group isomorphic with $S_N$ . Group theory and graphical techniques are then used to solve for the wave function exactly and analytically at each order.  Recently this formalism has been used to obtain the first-order, fully-interacting wave function for a system of harmonically-confined bosons interacting harmonically.  	In this paper, we report the first application of this $N$-body wave function to a system of $N$ fully-interacting bosons in three dimensions. We determine the density profile for a confined system of harmonically-interacting bosons.  Choosing this simple interaction is not necessary or even advantageous for our method, however this choice allows a direct comparison of our exact results through first order with exact results obtained in an independent solution. Our density profile through first-order in three dimensions is indistinguishable from the first-order exact result obtained independently and shows strong convergence to the exact result to all orders.
\end{abstract}

\maketitle

\section{Introduction}
The interacting quantum $N$-body problem becomes particularly challenging when $N$ is large. Unlike the corresponding classical problem
where the resources required to solve the problem scale as a polynomial in $N$\,, the resources needed to solve the quantum $N$-body
problem scale exponentially with $N$\,, frequently doubling for every particle added.\cite{liu:2007,montina2008} When interparticle interactions are
weak, the mean-field approximation may be used to avoid this exponential scaling with $N$\,. When interactions are larger, phenomenological models are often used. Typically the phenomenological models are only valid for a range of
interaction strengths, masses, \ldots etc., so more general first-principles approaches which tackle the quantum mechanical exponential $N$-scaling issue head on are needed.\cite{MSTTV,anderson:04} Such first-principles methods for confined, $N$-particle quantum systems include
coupled cluster methods (CCM)\cite{Cederbaum1,Cederbaum2}, the method of correlated basis
functions (CBF)\cite{fantoni:98:book,fabrocini:99}, and density functional theory\cite{Banerjee&Singh}. Of particular note in this regard are Monte Carlo methods\cite{MSTTV,anderson:04,Landau&Binder,holzmann:99,blume:01,nilsen:05,dubois:01,dubois:03,purwanto:05}.

We take a different approach and develop a non-numerical method which uses group theory and graphical techniques to tackle the $N$-scaling
problem (see Ref.~\onlinecite{wavefunction1harm} and the references therein).
This method circumvents the severity of the $N$-scaling problem by using a perturbation expansion about a maximally symmetric structure in large
dimensions which has a point group isomorphic to $S_N$\,. Group theory is then used to separate the $N$ scaling problem away from the interaction dynamics, allowing the $N$ scaling to be treated
as a straight mathematical issue. 
The perturbation expansion of the Hamiltonian is order-by-order invariant 
under the $S_N$ point group, yielding a problem at each perturbation order 
that can be solved, essentially exactly and analytically, using
group theory and graphical techniques. As part of this solution, small finite, $S_N$-invariant basis sets are used that 
are complete at each order, and as $N$ increases, the group theory and graphical
techniques ``hold their own'' with the result that the number of elements of this basis do not grow with $N$\,.
(The basis elements must remain invariant under the $N!$
operations of the $S_N$ group which puts increasing restrictions on the growth of the set as $N$ gets larger.)
The completeness of this basis at each order was established in
Ref.~\onlinecite{wavefunction1harm}. Since the elements of this basis, called binary invariants, are themselves invariant under
the maximal point group symmetry, the invariance of the Hamiltonian at each order is naturally ensured by expressing it in terms
of this relatively small basis (seven elements at lowest order in the wave function, twenty-three elements at next order).

Applying this approach at lowest order, we have previously derived beyond-mean-field energies\cite{energy,GFpaper}, frequencies\cite{energy},
normal mode coordinates\cite{normalmode}, wave functions\cite{normalmode} and density profiles\cite{density0}
for general isotropic, interacting confined quantum systems. More recently, in a major development of the method presented Refs.~\onlinecite{wavefunction1harm} and \onlinecite{wavefunction1}, we have extended this analysis to first order in the wave function, which required the development of new
techniques to handle, exactly and analytically, the complexity of coupling the many normal coordinates of the lowest-order solution. This basic approach, developed at first
order, is general enough to be extended to higher orders in the perturbation expansion.
In Ref.~\onlinecite{wavefunction1harm}, the general theory of Ref.~\onlinecite{wavefunction1}, which derives the
wave function through first order for an arbitrary isotropic system, was tested
on the exactly soluble system of $N$ harmonically-interacting particles under harmonic confinement. When this wave
function was compared to the exact analytic wave function obtained in an independent solution and then expanded analytically through first order, exact agreement was found, confirming this general theory for a fully interacting $N$-body system in three dimensions.\cite{wavefunction1harm}

In the present paper, we test this general, fully interacting wave function of Ref.~\onlinecite{wavefunction1}, exact through first order, by deriving
a property -- namely the density profile which is directly observable in the laboratory for a Bose-Einsten condensate. We test it on the same exactly soluble system of $N$ harmonically-interacting particles under harmonic confinement. Our density profile through first-order
evaluated at $D=3$ is indistinguishable from the $D=3$ first-order result from the independent solution and shows strong convergence to the exact $D=3$ result to all orders.

This derivation of the density profile through first order builds upon previous isotropic work. The isotropic, lowest-order ground state wave function
was derived in Ref.~\onlinecite{normalmode} and the corresponding lowest-order ground state density profile was derived
from this in Ref.~\onlinecite{density0}.
The isotropic, {\em first-order} ground state wave function was derived in Ref.~\onlinecite{wavefunction1}, and checked in
Ref.~\onlinecite{wavefunction1harm}. This work is reviewed in Section~\ref{sec:GSW_Review}, while  Appendix~\ref{app:Bin} contains
the briefest review of binary invariants. The derivation of the density profile through first order
from the general wave function through first order is found in Section~\ref{sec:Deriv_Dens}. Section~\ref{sec:test} presents the application and results.
Appendix~\ref{app:dens} presents the exact independent solution for the density profile through first order from the full $D$-dimensional density profile for this system, which is also derived independently in Appendix~\ref{app:dens} from the full $D$-dimensional
wave function for the harmonically interacting system discussed in Ref.~\onlinecite{wavefunction1harm}. Section~\ref{sec:Concl} is the summary and conclusions section.

\section{The Ground-State Wave Function}
\label{sec:GSW_Review}
\subsection{Lowest-Order Wave Function}
The zeroth-order Hamiltonian is that of a multi-dimensional
harmonic oscillator (see Eq.~(20) of Ref.~\onlinecite{wavefunction1}, or Eq.~(4) of Ref.~\onlinecite{wavefunction1harm}). Thus upon transformation to the normal
modes of the system, the wave function is a product of $P = N(N+1)/2$ one-dimensional harmonic
oscillator wave functions\cite{normalmode,wavefunction1}.
\begin{equation}
\Phi_0({\mathbf{q'}}) = \prod_{\nu=1}^P
\phi_{n_{\nu}}\left(  \sqrt{\bar{\omega}_\nu} 
\, q^\prime_{\nu} \right) \,, \label{eq:psiintro}
\end{equation}
where $\bar{\omega}_\nu$ is the frequency of normal mode $q^\prime_{\nu}$\,,
and $n_{\nu}$ is the oscillator quantum number, $0 \leq n_\nu
< \infty$, which counts the number of quanta in normal mode $\nu$.

\subsection{Normal Modes and Symmetry Coordinates}
The transformation to normal modes would appear to be a formidable proposition since if $N$ is in the millions,
the number of normal modes $P$ is of the order $10^{12}$ or larger. However, the $D \rightarrow \infty$ structure is maximally symmetric (see Ref.~\onlinecite{GFpaper}),  and it is this maximal point group symmetry which allows the normal modes to be derived (see Ref.~\onlinecite{normalmode}).

Under this $S_N$ point group symmetry, the normal modes transform under the $[N]$\,,
$[N-1, \hspace{1ex} 1]$\,, and $[N-2, \hspace{1ex} 2]$ irreducible representations. Two normal modes
transform under two one-dimensional $[N]$ irreducible representations (irreps.) of $S_N$\,,
$2(N-1)$ normal modes transform under two $(N-1)$-dimensional $[N-1, \hspace{1ex} 1]$ irreps.,
and $N(N-3)/2$ normal modes transform under an $N(N-3)/2$ dimensional $[N-2, \hspace{1ex} 2]$ irrep.\,.

The normal modes may be written as
\begin{equation} \label{eq:qnpfullexp}
{\bm{q}'}_\pm^\alpha = c_\pm^{\alpha} \left(
\cos{\theta^\alpha_\pm} \, [{\bm{S}}_{\bar{\bm{r}}'}^{\alpha}]_\xi
\, + \, \sin{\theta^\alpha_\pm} \,
[{\bm{S}}_{\overline{\bm{\gamma}}'}^{\alpha}]_\xi \right)
\end{equation}
for the $\alpha=[N]$ and $[N-1, \hspace{1ex} 1]$ sectors, where the $\pm$ denote the two normal mode vectors
for each $\alpha$\,, and

\begin{equation} \label{eq:qnm2fullexp}
{\bm{q}'}^{[N-2, \hspace{1ex} 2]} = c^{[N-2, \hspace{1ex} 2]}
{\bm{S}}_{\overline{\bm{\gamma}}'}^{[N-2, \hspace{1ex} 2]} \,,
\end{equation}
where the symmetry coordinates $[{\bm{S}}_{X'}^{\alpha}]_\xi$\,(defined in Ref.~\onlinecite{normalmode}) may be written
\begin{equation} \label{eq:S}
\renewcommand{\arraystretch}{1.5}
\begin{array}{@{}r@{\hspace{0.5ex}}c@{}l@{}}
{\mathbf{S}}_{r'}^{[N]} & = & {\displaystyle \frac{1}{\sqrt{N}} \,
\sum_{k=1}^N \overline{r}'_k \,,} \hspace{2em}
{\mathbf{S}}_{\gamma'}^{[N]} = {\displaystyle
\sqrt{\frac{2}{N(N-1)}} \,\,\, \sum_{l=2}^N \sum_{k =1}^{l-1}
\overline{\gamma}'_{kl} \,,}  \hspace{2em}
\\
\protect[{\mathbf{S}}_{r'}^{[N-1,1]}\protect]_i &=& {\displaystyle
\frac{1}{\sqrt{i(i+1)}} \left( \sum_{k=1}^i \overline{r}'_k - i
\overline{r}'_{i+1} \right)\,, 
} \\
\protect[{\mathbf{S}}_{\gamma'}^{[N-1,1]}\protect]_i & = &
{\displaystyle \frac{1}{\sqrt{i(i+1)(N-2)}} \, \left( \left[
\sum_{l = 2}^i \, \sum_{k=1}^{l-1} \hspace{-1ex}
\overline{\gamma}'_{kl} + \sum_{k = 1}^i \, \sum_{l=k+1}^{N}
\hspace{-1ex} \overline{\gamma}'_{kl} \right] - i \left[
\sum_{k=1}^i \overline{\gamma}'_{k,\,i+1} + \sum_{l=i+2}^N
\hspace{-0.5ex}
\overline{\gamma}'_{i+1,\,l} \right] \right) \,,} \\
\multicolumn{3}{l}{\hspace{2ex} \mbox{where} \hspace{2ex} 1 \leq i
\leq N-1 \,, \hspace{2ex} \mbox{and} } \\
\protect[{\mathbf{S}}_{\gamma'}^{[N-2,2]}\protect]_{ij} & = &
{\displaystyle \frac{1}{\sqrt{i(i+1)(j-3)(j-2)}} \, \left(
\vphantom{\sum_{k=1}^{[j'-1, i]_{min}} \hspace{-2ex}
\overline{\gamma}'_{kj'}} \right. }
\begin{array}[t]{@{}c@{}}
\displaystyle{ \hspace{-5ex} \sum_{j'=2}^{j-1} \sum_{k=1}^{[j'-1,
i]_{min}} \hspace{-2ex} \overline{\gamma}'_{kj'} +
\sum_{k=1}^{i-1} \sum_{j'=k+1}^i
\overline{\gamma}'_{kj'} - (j-3) \sum_{k=1}^i \overline{\gamma}'_{kj} - } \\
\left. \displaystyle{ - i \, \left( \sum_{k=1}^{i}
\overline{\gamma}'_{k,(i+1)} + \sum_{j'=i+2}^{j-1}
\overline{\gamma}'_{(i+1),j'} \right) + i (j-3)
\overline{\gamma}'_{(i+1),j} } \right)  \,,
\end{array}
\\
\multicolumn{3}{l}{\hspace{2ex} \mbox{where} \hspace{2ex} 1 \leq i
\leq j-3 \hspace{2ex} \mbox{and} \hspace{2ex} i+3 \leq j \leq N
\,. }
\end{array}
\renewcommand{\arraystretch}{1}
\end{equation}

All of the normal modes belonging to a normal mode vector transforming irreducibly under $S_N$ have the same collective motion frequency, so that instead of $N(N+1)/2$ possible distinct frequencies we only have {\em five} distinct
frequencies. Thus writing the wave function in terms of the abbreviated $S_N$ irrep.\ labels
$\mathbf{0} = [N]$\,, $\mathbf{1} = [N-1, \hspace{1ex} 1]$\,, and $\mathbf{2} = [N-2, \hspace{1ex} 2]$\,,
we have
\begin{equation}
  \label{eq:wavefunct} \Phi_0 (\mathbf{q'}) = 
  \prod_{\mu =\{\mathbf{0}^{\pm}, \hspace{0.5ex} \mathbf{1}^{\pm},
  \hspace{0.5ex} \mathbf{2}\}} \hspace{0.25em} \prod_{\xi = 1}^{d_{\mu}}
  \hspace{0.25em} \phi_{n (\mu, \xi)} \left( \sqrt{\bar{\omega}_{\mu}}\,[q'_{\mu}]_{\xi} \right)\,,
\end{equation}
where $\phi_{n (\mu, \xi)} \left(
\sqrt{\bar{\omega}_{\mu}}\,
[{q}'_{\mu}]_{\xi} \right)$ is a one-dimensional harmonic-oscillator
wave function of frequency $\bar{\omega}_{\mu}$ and $n (\mu, \xi) _{}$ is the
oscillator quantum number, $0 \leq n (\mu, \xi) < \infty${\hspace{0.25em}},
which counts the number of quanta in each normal mode. The index $\mu$
labels the manifold of normal modes with the same frequency
$\bar{\omega}_{\mu}$ while degeneracy of the $\mu$th normal mode is denoted $d_{\mu} = 1${\hspace{0.25em}}, $N - {\nobreak} 1$
or $N (N - {\nobreak} 3) / 2$ for $\mu = \mathbf{0}^{\pm}${\hspace{0.25em}},
$\mathbf{1}^{\pm}$ or $\mathbf{2}$, respectively.

\subsection{Lowest-Order Ground-State Wave Function}

The lowest-order wave function ${}_g \hspace{-0.25em} \Phi_0 (\mathbf{q'})$ for the ground state is given by
Eq.~(\ref{eq:wavefunct}) with all of the $\text{$n_\nu$}_{}$ set
equal to zero, i.e.
 \begin{equation}
     \label{eq:groundWF}{}_g \hspace{-0.25em} \Phi_0 (\mathbf{q'}) = \prod_{\nu = 1}^{P}
     \hspace{0.25em} \hspace{0.25em} \phi_0 
     \left( \sqrt{\bar{\omega}_{\nu}} \,
     \hspace{0.25em} q'_{\nu} \right),
   \end{equation}
where $\phi_0$ is the wave function of a single harmonic-oscillator,
   \begin{equation}
     \label{eqphi0} \phi_0 
     \left(
     \sqrt{\bar{\omega}_{\nu}} \,q'_{\nu} \right) = \left(
     \frac{\bar{\omega}_{\nu}}{\pi} \right)^{\frac{1}{4}}\exp \left( - \frac{1}{2}  \bar{\omega}_{\nu}\,{q'_{\nu}}^2 \right)\,.
   \end{equation}

\subsection{First-Order Wave Function}
Previous applications of dimensional perturbation theory went to very high order in the asymptotic $1/D$ expansion. For systems with a large number of degrees of freedom, the derivation of these high order terms can be computationally prohibitive and subject to numerical difficulties. In Ref.~\onlinecite{matrix_method} Dunn \emph{et. al.} present an algorithm by which these corrections may be derived exactly using tensor algebra. Using this formalism, the first-order correction to the lowest-order wave function is derived in Ref.~\onlinecite{wavefunction1}. Writing the wave function through
first-order as
\begin{equation}\label{eq:Phi1hat}
 \Phi_1 (\mathbf{q'}) = (1 + \delta^{\frac{1}{2}} \hat{\Delta})
   \Phi_0 (\mathbf{q'})\,,
\end{equation}
then $\hat{\Delta}$ satisfies the commutator equation
\begin{equation}\label{eq:deltahatcommute}
 [\hat{\Delta},\bar{H}_0]\Phi_0=\bar{H}_1\Phi_0.
\end{equation}
To solve this equation, it is helpful to note that since $\Phi_0 (\mathbf{q'})$ is a Gaussian function, the
derivatives in $\bar{H}_1$ and $\bar{H}_0$ written in normal coordinates "bring down"
normal coordinates from the exponent so that
$\bar{H}_1$ effectively becomes a 3rd-order polynomial of only odd powers in $\mathbf{q'}$.

\subsubsection{Evaluation of Derivatives in $\hat{H}_1\,\Phi_0$}\label{sec:h1deriv}
We evaluate the derivatives implicit in Eq.~(\ref{eq:deltahatcommute}), noting the Gaussian form of $\Phi_0 (\mathbf{q'})$, to reduce the operator equation to a polynomial equation in
${q}'_{\nu}$\,,
\begin{equation}
  \partial_{{q'}_{\nu}} \Phi_0 (\mathbf{q'}) =
  - \bar{\omega}_{\nu}{{q'}}_{\nu}\Phi_0 (\mathbf{q'})
\end{equation}
\begin{equation}
\partial^2_{q'_\nu}\Phi_0(\mathbf{q'}) =\left(-\bar{\omega}_\nu+\bar{\omega}_\nu^2({q'_\nu})^2\right)\Phi_0(\mathbf{q'}) \,.
\end{equation}

Therefore with the substitutions
\begin{eqnarray}
\partial_{{q'}_{\nu}}
&\rightarrow& - \bar{\omega}_{\nu}{{q'}}_{\nu}\\
\partial_{{q'}_{\nu_i}}\partial_{{q'}_{\nu_j}}&\rightarrow&\bar{\omega}_{\nu_i}\bar{\omega}_{\nu_j}{{q'}}_{\nu_i}{{q'}}_{\nu_j}-\delta_{ij}\,\bar{\omega}_{\nu_i}\,,
\end{eqnarray}
the action of $\bar{H}_1$ on $\Phi_0 (\mathbf{q'})$ becomes
equivalent to the action of a 3rd-order polynomial $\left(\bar{H}_1\right)_\textrm{eff}$ on $\Phi_0 (\mathbf{q'})$:
\begin{equation}
 \bar{H}_1 \Phi_0 (\mathbf{q'}) = \left(\bar{H}_1\right)_\textrm{eff} \Phi_0 (\mathbf{q'})
\end{equation}
where
\begin{eqnarray} \lefteqn{\left(\bar{H}_1\right)_\textrm{eff} = \sum_{\nu_1,\nu_2,\nu_3}\left(-\frac{1}{2} \left[\stacklr{1}{3}{G}{}{V}\right]_{\nu_1, \nu_2, \nu_3} 
   \bar{\omega}_{\nu_2}  \bar{\omega}_{\nu_3} + \frac{1}{3!}
   \left[\stacklr{1}{3}{F}{}{V}\right]_{\nu_1, \nu_2, \nu_3}\right) {q'}_{\nu_1}
   {q'}_{\nu_2} {q'}_{\nu_3}
   }
\nonumber\\&& + \sum_{\nu_1}\left(\frac{1}{2} \sum_{\nu_2} \left[\stacklr{1}{3}{G}{}{V}\right]_{\nu_1, \nu_2, \nu_2} 
   \bar{\omega}_{\nu_2}+\frac{1}{2} \left[\stacklr{1}{1}{G}{}{V}\right]_{\nu_1}  \bar{\omega}_{\nu_1} + \left[\stacklr{1}{1}{F}{}{V}\right]_{\nu_1} \right) {q'}_{\nu_1} .
\end{eqnarray}
%

We define the $(4\times4\times4)$ tensor $\tau^{H_1}_{\mu_1, \mu_2, \mu_3}$ and the length-$4$ column vector $\tau^{H_1}_{\mu_1}$ so that the above equation may be written in terms of Clebsch-Gordon coefficients of $S_N$\,, $C_{\xi_1, \xi_2, \xi_3}^{\mu_1
 \mu_2 \mu_3 \mathcal{R}}$\,, (see Section V.B of Ref.~\onlinecite{wavefunction1}) and their coefficient tensors
\begin{eqnarray}
  \tau^{H_1}_{\mu_1, \mu_2, \mu_3, \mathcal{R}}  
&=&
 -\frac{1}{2} \stacklr{1}{3}{\tau}{G}{\mu_1, \mu_2 ,\mu_3,\mathcal{R}}
  \bar{\omega}_{\mu_2}  \bar{\omega}_{\mu_3} + \frac{1}{3!}
  \stacklr{1}{3}{\tau}{F}{\mu_1, \mu_2 ,\mu_3,\mathcal{R}}
\nonumber\\
\tau^{H_1}_{\mu_1} &=& \frac{1}{2}\sum_{\mu_2}d_{\mu_2} \stacklr{1}{3}{\tau}{G}{\mu_1,\mu_2,\mu_2, \Rmn{1} } \, \bar{\omega}_{\mu_2}+\frac{1}{2} \stackleft{1}{1}\tau_{\mu_1}^G  \bar{\omega}_{\mu_1} + \stackleft{1}{1}\tau_{\mu_1}^F \,.
\end{eqnarray}
Therefore, the polynomial $\left(\bar{H}_1\right)_\textrm{eff}$ may be written in the following compact form:
\begin{eqnarray}\label{eq:H1polynomial}
  \left(\bar{H}_1\right)_\textrm{eff} &=& \underset{\mu_1, \mu_2, \mu_3,\mathcal{R}}{\sum}\tau^{H_1}_{\mu_1, \mu_2,
  \mu_3, \mathcal{R}}  
\underset{\xi_1, \xi_2, \xi_3}{\sum} C_{\xi_1, \xi_2, \xi_3}^{\mu_1
 \mu_2 \mu_3 \mathcal{R}} [{q'}_{\mu_1}]^{}_{\xi_1} [{q'}_{\mu_2}]^{}_{\xi_2}
 [{q'}_{\mu_3}]^{}_{\xi_3} +\nonumber\\&&
 + \sum_{\mu=\{\mathbf{0+}, \mathbf{0-} \}}\tau^{H_1}_{\mu} {q'}_{\mu}\,.
\end{eqnarray}

\subsubsection{Derivation of the cubic $\Delta$}
From Eqs.~(\ref{eq:deltahatcommute}) and (\ref{eq:H1polynomial}) (in Sec.~\ref{sec:h1deriv}), we obtain the polynomial equation

\begin{equation}\label{eq:deltacommute}
 [\Delta,H_0]\Phi_0=\left(\bar{H}_1\right)_\textrm{eff}\Phi_0.
\end{equation}
Solving this equation for the polynomial $\Delta$, we obtain
\begin{equation}\label{eq:deltadef}
\Delta=\sum_{\mu_1,\mu_2,\mu_3,\mathcal{R}}\sum_{\xi_1,\xi_2,\xi_3}\left(
\stacklr{1}{3}{\tau}{\Delta}{\mu_1,\mu_2,\mu_3, \mathcal{R}}C^{\mu_1\mu_2\mu_3,\mathcal{R}}_{\xi_1,\xi_2,\xi_3}
\right)
[{q'}_{\mu_1}]_{\xi_1}[{q'}_{\mu_2}]_{\xi_2}[{q'}_{\mu_3}]_{\xi_3} + \sum_{\mu=\{\mathbf{0+}, \mathbf{0-} \}}\stacklr{1}{1}{\tau}{\Delta}{\mu}{q'}_{\mu} \,,
\end{equation}
where
\begin{eqnarray}\label{eq:tauDelta}
\stackleft{1}{3}\tau^\Delta_{\mu_1,\mu_2,\mu_3,k}
&=&
\frac{-\stackleft{1}{3}\tau^{H_1}_{\mu_1,\mu_2,\mu_3,k}}{\bar{\omega}_{\mu_1}+\bar{\omega}_{\mu_2}+\bar{\omega}_{\mu_3}}
 \\
 \stackleft{1}{1}{\tau}^{\Delta}_{\mathbf{0}\pm}
 &=&
 \frac{1}{\bar{\omega}_{\mathbf{0}\pm}}
\left(-\stacklr{1}{1}{\tau}{H_1}{\mathbf{0}\pm}
\right.
\\&&\left.+\sum_{\mu}d_\mu
\left(
\stacklr{1}{3}{\tau}{\Delta}{\mathbf{0}\pm\mu\mu}
+\stacklr{1}{3}{\tau}{\Delta}{\mu\mathbf{0}\pm\mu}
+\stacklr{1}{3}{\tau}{\Delta}{\mu\mu\mathbf{0}\pm}
\right)
\right)\,.
\end{eqnarray}

Therefore, the first-order manybody wave function is obtained by multiplying the lowest-order wave function by $\Delta$, a polynomial in $\mathbf{q'}$ given by Eqs~(\ref{eq:deltadef}) and (\ref{eq:tauDelta}):
\begin{equation}\label{eq:Phi1}
 \Phi_1 (\mathbf{q'}) = (1 + \delta^{\frac{1}{2}} \Delta)
   \Phi_0 (\mathbf{q'}).
\end{equation}

\section{Derivation of the Density Profile Through First Order}
\label{sec:Deriv_Dens}
\subsection{Recap: The Lowest-Order Density Profile}
In Ref.~\onlinecite{density0}, we derived the Jacobian-weighted density profile at lowest order by integrating over many of the degrees of freedom of the wave function, and transforming that integral from normal to internal coordinates. We arrived at (Eq.~(68) of Ref.~\onlinecite{density0})
\begin{eqnarray} \label{eq:Nrresult}
\hspace{-3ex} S(D) \, \mathcal{N}_0(r) = N \sqrt{\frac{D}{\kappa^2(D)}\frac{R}{\pi}  }\exp{\left(- R\,\left(r\frac{\,\sqrt{D}}{\kappa(D)} -
\sqrt{D}\,\bar{r}_{\infty}\right)^2 \right) } \,.
\end{eqnarray}
The factor $S(D)$ is the $D$-dimensional solid
angle\cite{avery:91},
\begin{equation}
S(D)=\frac{2 \,\, \pi^{\frac{D}{2}}}{\Gamma(\frac{D}{2})} \,,
\end{equation}
and the quantity $R$ (a number) is defined in Eq.~(\ref{eq:Rdef}) below.

Notice that the lowest-order Jacobian-weighted density profile is a Gaussian (normalized to $N$) centered around $r=\kappa(D) \, \bar{r}_{\infty}$\,, the
 $D\rightarrow\infty$
configuration radius in oscillator units (see Eqs.~(9) and
(13) of Ref.~\onlinecite{density0}). The form of this Gaussian function is flexible in the sense that its shape depends on the values of two quantities, $R$ and $\bar{r}_\infty$\,. However, this lowest-order Jacobian-weighted density profile is still limited to a symmetric shape about $r=\kappa(D) \, \bar{r}_{\infty}$\,. The first-order corrections will add further flexibility by allowing for asymmetry.

\subsection{First-Order Corrections}
The derivation of the first-order density profile is similar to that of the lowest-order density profile in that the same transformations are used to perform a change of coordinates. Integrals over the normal coordinates ${\bm{q}'}_{{\bf
0}^+}$\,, ${\bm{q}'}_{{\bf 0}^-}$\,, $[{\bm{q}'}_{{\bf
1}^+}]_{N-1}$\,, and $[{\bm{q}'}_{{\bf 1}^-}]_{N-1}$\, are transformed to $\bar{r}'_{N}$\,, $\bar{r}'_{S}$\,,
${\bm{S}}_{\overline{\bm{\gamma}}'}^{[N]}$\,, and
$[{\bm{S}}_{\overline{\bm{\gamma}}'}^{[N-1, \hspace{1ex}
1]}]_{(N-1)}$\,, where
\begin{equation} \label{eq:rs}
\bar{r}'_{S} = \sum_{i=1}^{N-1} \bar{r}'_{i} \,.
\end{equation}

The first-order density profile is derived from the first-order wave function in a similar way to lowest order derivation.
In Ref.~\onlinecite{density0} we showed that we can write
\begin{equation}\label{eq:Nsetup}
\renewcommand{\arraystretch}{1.5}
\begin{array}[b]{@{}l@{}}
{\displaystyle S(D) \, N(r) } \\
 {\displaystyle =  N \int_{-\infty}^\infty \hspace{-0.5ex} \cdots \int_{-\infty}^\infty
\hspace{-1.3ex} \delta_f(r-r_N) \,
[_g\!\left[\Phi({\mathbf{\bar{y}'}})\right]^2 \hspace{-2ex} \prod_{\mu= {\bf
0}^\pm, {\bf 1}^\pm, {\bf 2}} \, \prod_{\xi=1}^{d_{\mu}}
d[q'_\mu]_\xi \,.}
\end{array}
\renewcommand{\arraystretch}{1}
\end{equation}
By simply substituting $\left[{}_g\Phi_1 ( \mathbf{\bar{q}'})\right]^2=(1+\delta^{1/2} \Delta)^2\left[{}_g\Phi_0 ( \mathbf{\bar{q}'})\right]^2$ for $\left[{}_g\Phi ( \mathbf{\bar{q}'})\right]^2$ in Eq.~(\ref{eq:Nsetup}) we obtain the first-order density profile as
\begin{eqnarray}\label{eq:N1throughDelta2}
\mathcal{N}_1 (r) &=&  \frac{N}{S(D)} \int_{-\infty}^{\infty}  \hspace{-0.5ex} \cdots \int_{- \infty}^{\infty} 
\prod_{\mu = \mathbf{0}^{\pm},\mathbf{1}^{\pm}, \mathbf{2}}  \hspace{0.25em}
\prod_{\xi = 1}^{d_{\mu}}d [ {q}'^{\mu}]_{\xi} 
\nonumber\\&&
\times \delta_f (r - r_N)\left(1+2 \delta^{1 / 2} \Delta  + \delta \Delta^2\right)\left[{}_g\Phi_0 ( \mathbf{\bar{q}'})\right]^2\,,
\end{eqnarray}
where $\delta_f(r-r_i)$ is the Dirac delta function (differentiated from the inverse dimension, $\delta$\,, by the subscript $f$\,).
Thus to first order in $\delta^{1/2}$ the Jacobian-weighted density profile is

\begin{eqnarray}  \label{eq:N1throughDelta}
\mathcal{N}_1 (r) &=&\mathcal{N}_0 (r)+\frac{N}{S(D)}  \linebreak 2 \delta^{1 / 2} \int_{-
  \infty}^{\infty}  \hspace{-0.5ex} \cdots \int_{- \infty}^{\infty} \prod_{\mu = \mathbf{0}^{\pm},
  \mathbf{1}^{\pm}, \mathbf{2}}  \hspace{0.25em} \prod_{\xi = 1}^{d_{\mu}}
  d [ {q}'_{\mu}]_{\xi} 
\nonumber\\&&
\times  \delta_f (r - r_N) \Delta\hspace{.25em}\left[{}_g
  \Phi_0 ( \mathbf{\bar{q}'})\right]^2 \,.
\end{eqnarray}
Substituting $\Delta$ from Eq.~(\ref{eq:deltadef}), we obtain
\begin{eqnarray}  
\lefteqn{\mathcal{N}_1 (r) =\mathcal{N}_0 (r)+ \linebreak \frac{N}{S(D)}2 \delta^{1 / 2} \int_{-
  \infty}^{\infty}  \hspace{-0.5ex} \cdots \int_{- \infty}^{\infty} \prod_{\mu = \mathbf{0}^{\pm},
  \mathbf{1}^{\pm}, \mathbf{2}}  \hspace{0.25em} 
\prod_{\xi = 1}^{d_{\mu}}d[{q}'_{\mu}]_{\xi} \, \delta_f (r - r_N) 
}
\nonumber\\&&
\times 
\left(\underset{\mu_1,\mu_2,\mu_3,\mathcal{R}}{\sum}\;\underset{\xi_1,\xi_2,\xi_3}{\sum}
\stacklr{1}{3}{\tau}{\Delta}{\mu_1,\mu_2,\mu_3,\mathcal{R}}C^{\mu_1\mu_2\mu_3 ,\mathcal{R}}_{\xi_1,\xi_2,\xi_3}[{q'}_{\mu_1}]_{\xi_1}[{q'}_{\mu_2}]_{\xi_2}[{q'}_{\mu_3}]_{\xi_3} +
\sum_{\mu=\{\mathbf{0+}, \mathbf{0-} \}}\stacklr{1}{1}{\tau}{\Delta}{\mu}{q'}_{\mu}
\right)
\nonumber\\&&
\times 
\left[{}_g  \Phi_0 ( \mathbf{\bar{q}'})\right]^2 \,, \label{eq:N11}
\end{eqnarray}
where $\mu_1$\,, $\mu_2$\,, and $\mu_3$ each run over $\mathbf{0}^{+}${\hspace{0.25em}},
$\mathbf{0}^{-}${\hspace{0.25em}}, $\mathbf{1}^{+}$
$\mathbf{1}^{-}$\,, and $\mathbf{2}$\,, and $1 \leq \xi_1\,,\hspace{1ex}\xi_2\,,\hspace{1ex}\xi_3 \leq d_{\mu}$ where $d_{\mu} = 1${\hspace{0.25em}}, $N - {\nobreak} 1$
and $N (N - {\nobreak} 3) / 2$ for $\mu = \mathbf{0}^{\pm}$\,, $\mathbf{1}^{\pm}$\,, and $\mathbf{2}$ respectively. When $\mu = \mathbf{2}$\,, $\Rmn{1} \leq \mathcal{R} \leq \Rmn{2}$\,; otherwise $\mathcal{R} = \Rmn{1}$\,.

\subsubsection{Normal Coordinate Integrals}
We simplify Eq.~(\ref{eq:N11})
by defining the $(P\times P\times P)$ tensor ${}_3M_{\nu_1,\nu_2,\nu_3}$ of integrals and the length $P$ column vector ${}_1M_{\nu}$ of integrals
as special cases of the rank $n$ tensor of integrals
\begin{equation} \label{eq:Mdef}
 {}_nM_{\nu_1,\nu_2, \ldots \nu_n} =
\int_{-
  \infty}^{\infty}  \hspace{-0.5ex} \cdots \int_{- \infty}^{\infty} \prod_{\mu = \mathbf{0}^{\pm},
  \mathbf{1}^{\pm}, \mathbf{2}}  \hspace{0.25em} 
\prod_{\xi = 1}^{d_{\mu}}d[{q}'_{\mu}]_{\xi} \, \delta_f (r - r_N){q'}_{\nu_1}{q'}_{\nu_2} \cdots {q'}_{\nu_n}
\,\,  \left[{}_g
  \Phi_0 ( \mathbf{\bar{q}'})\right]^2 \,.
\end{equation}
Indexing ${}_3M$ and ${}_1M$ by $(\mu_i,\xi_i)$ rather than $\nu_i$ (See Ref.~\onlinecite{JMPEPAPS}, Table~\Rmn{3}), we can write the density profile as a tensor contraction
\begin{eqnarray}  \label{eq:N1_pre_int}
\lefteqn{\mathcal{N}_1 (r) =\mathcal{N}_0 (r)+ }
\\&&\linebreak \frac{2 \delta^{1 / 2}\,N}{S(D)} 
\left(\underset{\mu_1,\mu_2,\mu_3,\mathcal{R}}{\sum}\;\underset{\xi_1,\xi_2,\xi_3}{\sum}
\stacklr{1}{3}{\tau}{\Delta}{\mu_1,\mu_2,\mu_3,\mathcal{R}}C^{\mu_1\mu_2\mu_3 ,\mathcal{R}}_{\xi_1,\xi_2,\xi_3}
{}_3 M^{\mu_1\mu_2\mu_3}_{\xi_1,\xi_2,\xi_3}
+
\sum_{\mu=\{\mathbf{0+}, \mathbf{0-} \}} \stacklr{1}{1}{\tau}{\Delta}{\mu}{}_{1}M^{\mu}_{1}
\right) \nonumber\,,
\end{eqnarray}
Each element of the $M$ tensor is an integral.

Due to the presence of the Dirac delta function in ${}_3M$ and ${}_1M$ we must treat integrals over
normal coordinates involving $r_N$ differently from those that don't involve $r_N$\,. As in Ref.~\onlinecite{density0} which derives the lowest-order
density profile $\mathcal{N}_0 (r)$\,, we note that $r_{N}$
appears in only four normal coordinates, i.e.\  in ${\bm{q}'}_{{\bf 0}^+}$\,, ${\bm{q}'}_{{\bf 0}^-}$\,,
$[{\bm{q}'}_{{\bf 1}^+}]_{d_{{\bf 1}^+}}$ and $[{\bm{q}'}_{{\bf
1}^-}]_{d_{{\bf 1}^-}}$ (see Eqs.~(\ref{eq:qnpfullexp}), (\ref{eq:qnm2fullexp}) and (\ref{eq:S})). The remaining $(N(N+1)/2 -4)$ integrals, the overwhelming majority of the integrals,
are easily evaluated using the integral identity
\begin{equation} \label{eq:q_int}
\int_{-\infty }^{\infty }
   \sqrt{\frac{\bar{\omega}_{\nu }}{\pi }} 
   \left(q'_{\nu }\right)^m
   e^{-\bar{\omega }_{\nu } \left(q'_{\nu }\right)^2} \, dq'_{\nu }
   = \frac{(1 + \exp{(i\,m\,\pi)})}{2 \sqrt{\pi} \, \bar{\omega }_{\nu }^{\frac{m}{2}} } \, \Gamma\left(\frac{m+1}{2} \right) \,,
\end{equation}
which is zero when $m$ is odd. Thus many terms are zeroed out; only those terms involving even powers
of the 
normal coordinates which are independent of $r_N$ contribute.

Thus from Eqs.~(\ref{eq:N1_pre_int}) and (\ref{eq:q_int}) we find
\begin{eqnarray}  
\lefteqn{\mathcal{N}_1 (r)  =  \mathcal{N}_0 (r)+ \linebreak \frac{2 \delta^{1 / 2}\,N}{S(D)} 
\left(\sum_{\mu'_1,\mu'_2,\mu'_3}
\stacklr{1}{3}{\tau}{\Delta}{\mu'_1,\mu'_2,\mu'_3,\Rmn{1}}C^{\mu'_1\mu'_2\mu'_3 ,\Rmn{1}}_{d_{\mu'_1 },d_{\mu'_2 },d_{\mu'_3 }}
{}_3 M^{\mu'_1\mu'_2\mu'_3}_{d_{\mu'_1 },d_{\mu'_2 },d_{\mu'_3 }} \vphantom{\frac{1}{2 \bar{\omega}_{\bar{\mu}} }}
\right.}
  \nonumber \\ && +
\left. \sum_{\mu'}\left\{\sum_{\bar{\mu}} \frac{1}{2 \bar{\omega}_{\bar{\mu}} } ( \stacklr{1}{3}{\tau}{\Delta}{\bar{\mu},\bar{\mu},\mu',\Rmn{1}} + \stacklr{1}{3}{\tau}{\Delta}{\bar{\mu},\mu',\bar{\mu},\Rmn{1}}  + \stacklr{1}{3}{\tau}{\Delta}{\mu',\bar{\mu},\bar{\mu},\Rmn{1}}  )  \, C^{\bar{\mu}\bar{\mu}\mu' ,\Rmn{1}}_{\bar{\xi},\bar{\xi},d_{\mu' }} 
+
 \delta_{\mu',\,\bm{0}^\pm} \stacklr{1}{1}{\tau}{\Delta}{\mu'} \right\} \,
{}_{1}M^{\mu'}_{d_{\mu' }} 
\right)  \,, \nonumber \\
  \label{eq:N1tauCM}
\end{eqnarray}
where all primed indices range over ${\bf 0}^\pm$ and ${\bf 1}^\pm$\,, $\delta_{\mu',\,\bm{0}^\pm}$ equals one when $\mu' = \bm{0}^\pm$
but is zero otherwise, $\bar{\mu}$ ranges over ${\bf 1}^\pm$ and ${\bf 2}$\,,
and $1 \leq \bar{\xi} \leq d_{{\bf 1}^\pm } -1 = N-2$
when $\bar{\mu} = {\bf 1}^\pm$\,, or $1 \leq \bar{\xi} \leq d_{{\bf 2} } = N(N-3)/2$ when $\bar{\mu} = {\bf 2}$\,.

\subsubsection{Clebsch-Gordon tensor contractions}
The relevant Clebsch-Gordon elements and sums in the above equation are
\begin{equation}
\begin{array}{rclclcl}
C^{\bm{0}}_{1}{\, }^{\bm{0}}_{1}{\, }^{\bm{0}}_{1}{\, }^{,\Rmn{1}} & = & 1  &&&& \\
C^{\bm{1}}_{N-1}{\, }^{\bm{0}}_{1}{\, }^{\bm{0}}_{1}{\,}^{,\Rmn{1}} & = & C^{\bm{0}}_{1}{\, }^{\bm{1}}_{N-1}{\, }^{\bm{0}}_{1}{\,}^{,\Rmn{1}} & = &  C^{\bm{0}}_{1}{\, }^{\bm{0}}_{1}{\, }^{\bm{1}}_{N-1}{\,}^{,\Rmn{1}} & = &  0  \\
C^{\bm{1}}_{N-1}{\, }^{\bm{1}}_{N-1}{\, }^{\bm{0}}_{1}{\,}^{,\Rmn{1}} & = & C^{\bm{1}}_{N-1}{\, }^{\bm{0}}_{1}{\, }^{\bm{1}}_{N-1}{\,}^{,\Rmn{1}} & = &  C^{\bm{0}}_{1}{\, }^{\bm{1}}_{N-1}{\, }^{\bm{1}}_{N-1}{\,}^{,\Rmn{1}} & = &  1  \\[1ex]
C^{\bm{1}}_{N-1}{\, }^{\bm{1}}_{N-1}{\, }^{\bm{1}}_{N-1}{\,}^{,\Rmn{1}} & = & {\displaystyle \frac{- (N - 2)}{\sqrt{N (N - 1)}} }
 &&&&
\end{array}
\end{equation}
\begin{eqnarray}
\sum_{\bar{\xi}=1}^{N-2} C^{\bm{1}}_{\bar{\xi}}{\, }^{\bm{1}}_{\bar{\xi}}{\, }^{\bm{0}}_1{\, }^{,\Rmn{1}} & = & N-2 \\
\sum_{\bar{\xi}=1}^{N(N-3)/2} C^{\bm{2}}_{\bar{\xi}}{\, }^{\bm{2}}_{\bar{\xi}}{\, }^{\bm{0}}_1{\, }^{,\Rmn{1}} & = & 
\frac{N(N-3)}{2} \\
\sum_{\bar{\xi}=1}^{N-2} C^{\bm{1}}_{\bar{\xi}}{\, }^{\bm{1}}_{\bar{\xi}}{\, }^{\bm{1}}_{N-1}{\, }^{,\Rmn{1}} & = & \frac{N-2}{\sqrt{N(N-1)}} \\
\sum_{\bar{\xi}=1}^{N(N-3)/2} C^{\bm{2}}_{\bar{\xi}}{\, }^{\bm{2}}_{\bar{\xi}}{\, }^{\bm{1}}_{N-1}{\, }^{,\Rmn{1}} & = & 
0 \,.\label{s2s21nm1}
\end{eqnarray}
Equation~(\ref{s2s21nm1}) simply follows from the fact that since the $\bm{2}$ indices are saturated, there is nothing to couple the $\bm{1}$ index with to form a scalar $\bm{0}$ irrep.\,.

Equation~(\ref{eq:N1tauCM}) may be further simplified by defining two tensors 
which hold the coefficients of the elements of ${}_3 M^{\mu'_1\mu'_2\mu'_3}_{d_{\mu'_1 },d_{\mu'_2 },d_{\mu'_3 }}$ and ${}_{1}M^{\mu'}_{d_{\mu' }} $ tensors.
We define a $4\times4\times4$ tensor, ${}^3 E_{\mu'_1, \mu'_2, \mu'_3}$\,, to be the
non-zero coefficients of the elements of ${}_3 M^{\mu'_1\mu'_2\mu'_3}_{d_{\mu'_1 },d_{\mu'_2 },d_{\mu'_3 }}$ (which are cubic in $r_N$)
\begin{equation} \label{eq:E3}
 {}^3 E_{\mu'_1, \mu'_2, \mu'_3} = \stacklr{1}{3}{\tau}{\Delta}{\mu'_1,\mu'_2,\mu'_3,\Rmn{1}}C^{\mu'_1\mu'_2\mu'_3 ,\Rmn{1}}_{d_{\mu'_1 },d_{\mu'_2 },d_{\mu'_3 }} \,.
\end{equation}
We emphasize that the above equation is a simple elemental multiplication, no summation is implied. Using the above Clebsch-Gordon elements, we obtain
\begin{eqnarray}
  {}^3 E_{\mathbf{0} \pm, \mathbf{0} \pm, \mathbf{0} \pm} &=&
  {}^{} \tau^{\Delta}_{\mathbf{0} \pm, \mathbf{0} \pm, \mathbf{0}
  \pm} 
\label{eq:3Ea} \\
  {}^3 E_{\mathbf{1} \pm, \mathbf{0} \pm, \mathbf{0} \pm} &=&
  {}^3 E_{\mathbf{0} \pm, \mathbf{1} \pm, \mathbf{0} \pm} =
  {}^3 E_{\mathbf{0} \pm, \mathbf{0} \pm, \mathbf{1} \pm} = 0 
\label{eq:3Eb} \\
  {}^3 E_{\mathbf{1} \pm, \mathbf{1} \pm, \mathbf{0} \pm} &=&
  \tau^{\Delta}_{\mathbf{1} \pm, \mathbf{1} \pm, \mathbf{0}
  \pm}
\label{eq:3Ec} \\
  {}^3 E_{\mathbf{1} \pm, \mathbf{0} \pm, \mathbf{1} \pm} &=&
  \tau^{\Delta}_{\mathbf{1} \pm, \mathbf{0} \pm, \mathbf{1}
  \pm}
\label{eq:3Ed} \\
  {}^3 E_{\mathbf{0} \pm, \mathbf{1} \pm, \mathbf{1} \pm} &=&
  \tau^{\Delta}_{\mathbf{0} \pm, \mathbf{1} \pm, \mathbf{1}
  \pm}
\label{eq:3Ee} \\
  {}^3 E_{\mathbf{1} \pm, \mathbf{1} \pm, \mathbf{1} \pm} &=&
  \frac{- (N - 2)}{\sqrt{N (N - 1)}} {}^{}
  \tau^{\Delta}_{\mathbf{1} \pm, \mathbf{1} \pm, \mathbf{1} \pm} \,. \label{eq:3Ef}
\end{eqnarray}
In Eqs.~(\ref{eq:3Ea})--(\ref{eq:3Ef}), each $\pm$ associated with a sector $\mu$ is taken to be independent of the $\pm$ associated with the other two sectors.

We also define a length-four column vector ${}^1 E_{\mu' }$ to be the non-zero
coefficients of the elements of ${}_{1}M^{\mu'}_{d_{\mu' }}$ (which are linear in $r_N$).
\begin{equation} \label{eq:E1}
{}^1 E_{\mu' } = \sum_{\bar{\mu}}\frac{1}{2 \bar{\omega}_{\bar{\mu}} } \left( \stacklr{1}{3}{\tau}{\Delta}{\bar{\mu},\bar{\mu},\mu',\Rmn{1}} + \stacklr{1}{3}{\tau}{\Delta}{\bar{\mu},\mu',\bar{\mu},\Rmn{1}}  + \stacklr{1}{3}{\tau}{\Delta}{\mu',\bar{\mu},\bar{\mu},\Rmn{1}}  \right)  \sum_{\bar{\xi}}\, C^{\bar{\mu}\bar{\mu}\mu' ,\Rmn{1}}_{\bar{\xi},\bar{\xi},d_{\mu' }} + \delta_{\mu',\,\bm{0}^\pm} \stacklr{1}{1}{\tau}{\Delta}{\mu'} \,,
\end{equation}
i.e.\
\begin{eqnarray}
  {}^1 E_{\mathbf{0} \pm} 
&=& 
\frac{(N-2)}{2}\frac{1}{\omega_{\mathbf{1+}}}
\left(\tau^{\Delta}_{\mathbf{1+1+0\pm}}+\tau^{\Delta}_{\mathbf{1+0\pm1+}}+\tau^{\Delta}_{\mathbf{0\pm1+1+}}\right)
\nonumber\\&&
+
\frac{(N-2)}{2}\frac{1}{\omega_{\mathbf{1-}}}
\left(\tau^{\Delta}_{\mathbf{1-1-0\pm}}+\tau^{\Delta}_{\mathbf{1-0\pm1-}}+\tau^{\Delta}_{\mathbf{0\pm1-1-}}\right)
\nonumber\\&&
+
\frac{N(N-3)}{4}\frac{1}{\omega_{\mathbf{2}}}
\left(\tau^{\Delta}_{\mathbf{220\pm}}+\tau^{\Delta}_{\mathbf{20\pm2}}+\tau^{\Delta}_{\mathbf{0\pm22}}\right)+\stacklr{1}{1}{\tau}{\Delta}{\mathbf{0\pm}}
\\
  {}^1 E_{\mathbf{1} \pm} 
&=& 
\frac{N-2}{2\sqrt{N(N-1)}}\frac{1}{\omega_{\mathbf{1+}}}\left(\stacklr{1}{3}{\tau}{\Delta}{\mathbf{1+1+1\pm}}+\stacklr{1}{3}{\tau}{\Delta}{\mathbf{1+1\pm1+}}+\stacklr{1}{3}{\tau}{\Delta}{\mathbf{1\pm1+1+}}\right)
\nonumber\\&&
+
\frac{N-2}{2\sqrt{N(N-1)}}\frac{1}{\omega_{\mathbf{1-}}}\left(\stacklr{1}{3}{\tau}{\Delta}{\mathbf{1-1-1\pm}}+\stacklr{1}{3}{\tau}{\Delta}{\mathbf{1-1\pm1-}}+\stacklr{1}{3}{\tau}{\Delta}{\mathbf{1\pm1-1-}}\right) \,.
\end{eqnarray}
Using Eqs.~(\ref{eq:E3}) and (\ref{eq:E1}), Eq.~(\ref{eq:N1tauCM}) may be written in a simpler form as
\begin{equation}  \label{eq:N1EM}
\mathcal{N}_1 (r) = \mathcal{N}_0 (r)+ \linebreak \frac{2 \delta^{1 / 2}\,N}{S(D)} 
\left(\sum_{\mu'_1,\mu'_2,\mu'_3}
{}^3 E_{\mu'_1, \mu'_2, \mu'_3} \,\,
{}_3 M^{\mu'_1\mu'_2\mu'_3}_{d_{\mu'_1 },d_{\mu'_2 },d_{\mu'_3 }} \vphantom{\frac{1}{2 \bar{\omega}_{\bar{\mu}} }}  +
\sum_{\mu'}
{}^1 E_{\mu' }  \,
{}_{1}M^{\mu'}_{d_{\mu' }} 
\right)
\end{equation}

\subsubsection{Transformation of the integrals to symmetry coordinates}
The elements of the $M$ tensors are integrals over the normal coordinates. We use the $\bm{T}$ transformation of Ref.~\onlinecite{density0}, where
\begin{equation} \label{eq:qTa}
\left( \begin{array}{c} {\bm{q}'}^{{\bf 0}^+}
\\ {\bm{q}'}^{{\bf 0}^-} \\ \protect[{\bm{q}'}^{{\bf
1}^+}\protect]_{N-1} \\ \protect[{\bm{q}'}^{{\bf
1}^-}\protect]_{N-1} \end{array} \right) = \, \bm{T} \,\,
{\bm{a}'} \,,
\end{equation}
to write these $M$ tensors in terms of integrals over the coordinates of the four-element vector
\begin{equation}
\bm{a}' = \left( \begin{array}{c} \bar{r}'_{N} \\ \bm{b}' \end{array} \right) \,,
\end{equation}
where
\begin{equation} \label{eq:vector_b}
\bm{b}' = \left( \begin{array}{c} \bar{r}'_{S}
\\ {\bm{S}}_{\overline{\bm{\gamma}}'}^{[N]} \\ \protect[{\bm{S}}_{\overline{\bm{\gamma}}'}^{[N-1, \hspace{1ex}
1]}\protect]_{(N-1)} \end{array} \right) \,,
\end{equation}
i.e.\
\begin{equation} \label{eq:McalM}
{}_n M^{\mu'_1}_{d_{\mu'_1 }}{}^{\mu'_2}_{d_{\mu'_2 }}{ }^{\cdots \,\, \mu'_n}_{\cdots \,\, d_{\mu'_n }} =\sum_{i_1,i_2,\ldots,i_n} T_{\mu'_1 ,\, i_1}
 \, T_{\mu'_2 ,\, i_2} \, \cdots \, T_{\mu'_n ,\, i_n} \,\, {}_n  \mathcal{M}_{i_1\, i_2 \, \cdots \, i_n}\,,
\end{equation}
where $\bm{T}$ is given in Eqs.~(53), (54), and (55) of Ref.~\onlinecite{density0}. Applying the analysis of
Section~\Rmn{8} of Ref.~\onlinecite{density0} to the ${}_n  \mathcal{M}_{i_1\, i_2 \, \cdots \, i_n}$
tensor, we arrive at
\begin{equation} \label{eq:calMint}
\renewcommand{\arraystretch}{2}
\begin{array}[b]{@{}l@{}}
{}_n  \mathcal{M}_{i_1\, i_2 \, \cdots \, i_n} = \begin{array}[t]{@{}l@{}} {\displaystyle \frac{ J_T \sqrt{
\bar{\omega}_{{\bf 0}^+} \, \bar{\omega}_{{\bf 0}^-} \,
\bar{\omega}_{{\bf 1}^+} \, \bar{\omega}_{{\bf 1}^-} } }{\pi^2} \!
\int_{-\infty}^\infty \int_{-\infty}^\infty \int_{-\infty}^\infty
\int_{-\infty}^\infty \!\!\!\! \delta_f(r-a_1) } \\
{\displaystyle \hspace{2ex} \times \hspace{1ex} a_{i_1} a_{i_2} \cdots a_{i_n} \, \exp{\left(- K_0 \,
a_1^2 - 2 a_1 \bm{K}^T \bm{b}' -
\bm{b}'^T \bm{\mathcal{K}} \bm{b}' \right) } \, da_1 \,
d^3{\bm{b}'} \,, }
\end{array} \end{array}
\renewcommand{\arraystretch}{1}
\end{equation}
where $K_0$\,, $\bm{K}$\,, and $\bm{\mathcal{K}}$ are defined in Eqs.~(59), (60) and (61) respectively
of Ref.~\onlinecite{density0} and $J_T$ is the Jacobian of the transformation from
${\bm{q}'}_{{\bf 0}^+}$\,, ${\bm{q}'}_{{\bf 0}^-}$\,,
$[{\bm{q}'}_{{\bf 1}^+}]_{d_{{\bf 1}^+}}$ and $[{\bm{q}'}_{{\bf
1}^-}]_{d_{{\bf 1}^-}}$ to the
internal coordinates $\bm{a}'$\,, Eq.~(55) of Ref.~\onlinecite{density0}.

Substituting Eq.~(\ref{eq:McalM}) into Eq.~(\ref{eq:N1EM}) we obtain the density profile in terms of integrals over internal coordinates,
i.e.\
\begin{equation}  \label{eq:N1SigmaCalM}
\mathcal{N}_1 (r) = \mathcal{N}_0 (r)+ \linebreak \frac{2 \delta^{1 / 2}\,N}{S(D)} \left(\sum_{i_1,i_2,i_3}
{}_3\Xi_{i_1 \, i_2 \, i_3 } \,\,
{}_3  \mathcal{M}_{i_1\, i_2 \,  i_3} \vphantom{\frac{1}{2 \bar{\omega}_{\bar{\mu}} }}  +\sum_i
{}_1\Xi_{i}  \,
{}_1  \mathcal{M}_{i}\right) \,,
\end{equation}
where the (length 4) column vector ${}_1\Xi_{i}$ and the $(4\times4\times4)$ tensor ${}_3\Xi_{i,j,k}$ are
\begin{eqnarray}
{}_1\Xi_{i} & = & \sum_{\mu'}{}^1 E_{\mu' } \,\, T_{\mu' ,\, i} \\
{}_3\Xi_{i_1 \, i_2 \, i_3 } & = & \sum_{\mu'_1,\mu'_2,\mu'_3} {}^3 E_{\mu'_1, \mu'_2, \mu'_3}  \,\, T_{\mu'_1 ,\, i_1} \, T_{\mu'_2 ,\, i_2} \, T_{\mu'_3 ,\, i_3} \,.
\end{eqnarray}
%


\subsubsection{Evaluation of the integrals}
To perform the integrals of Eq.~(\ref{eq:calMint}) for
${}_3  \mathcal{M}_{i_1\, i_2 \,  i_3}$ and ${}_1  \mathcal{M}_{i}$ in Eq.~(\ref{eq:N1SigmaCalM}), we
define the four-component vector
\begin{equation}
\bm{V} = \left(  \begin{array}{c}  \frac{1}{2} \, K_0  \\ \bm{K}  \end{array} \right) \,.
\end{equation}
Eq.~(\ref{eq:calMint}) can be written as a series of derivatives of a term proportional to $\mathcal{N}_0(r)$
\begin{equation} \label{eq:Mderiv}
\renewcommand{\arraystretch}{2}
\begin{array}[t]{@{}l@{}}
{}_n  \mathcal{M}_{i_1\, i_2 \, \cdots \, i_n} = \begin{array}[t]{@{}l@{}} {\displaystyle \frac{ 
\sqrt{\det{\bm{\mathcal{K}}} } \, R}{\pi^2} }
\\  \hspace{2ex} \times {\displaystyle
\left( \frac{-1}{2 \bar{r}'}  \frac{\partial}{\partial V_{i_1}} \right) \left( \frac{-1}{2 \bar{r}'}  \frac{\partial}{\partial V_{i_2}} \right)
 \cdots \left( \frac{-1}{2 \bar{r}'}  \frac{\partial}{\partial V_{i_n}} \right)
\!
\int_{-\infty}^\infty \int_{-\infty}^\infty \int_{-\infty}^\infty
\int_{-\infty}^\infty \!\!\!\! \delta_f(r-a_1) } \\
{\displaystyle \hspace{2ex} \times \hspace{1ex}  \exp{\left(- K_0 \,
a_1^2 - 2 a_1 \bm{K}^T \bm{b}' -
\bm{b}'^T \bm{\mathcal{K}} \bm{b}' \right) } \, da_1 \,
d^3{\bm{b}'} \,, }
\end{array} \end{array}
\renewcommand{\arraystretch}{1}\,,
\end{equation}
where we have also used Eq.~(67) of Ref.~\onlinecite{density0},
\begin{equation}\label{eq:Rdef}
R = \frac{ \bar{\omega}_{{\bf 0}^+} \, \bar{\omega}_{{\bf 0}^-} \,
\bar{\omega}_{{\bf 1}^+} \, \bar{\omega}_{{\bf 1}^-} \, J_T^2 }{
\det{\bm{\mathcal{K}}}} =(K_0 - \bm{K}^T \bm{\mathcal{K}}^{-1}
\bm{K} ) \,.
\end{equation}
Upon using the integral identity
\begin{eqnarray} \label{eq:multidimgaussintegral}
\lefteqn{ \hspace{-10ex} \int_{-\infty}^\infty \cdots
\int_{-\infty}^\infty \exp{\left( - \bm{b}'^T \bm{\mathcal{K}} \bm{b}' -  2
a_1 \, \bm{K}^T \bm{b}' \right) } \, d^n{\bm{b}'} } \nonumber \\ &
\hspace{4ex} = & \frac{\pi^{\frac{n}{2}}}{\sqrt{\det{\bm{\mathcal{K}}}}}
\exp{\left( a_1^2 \, \bm{K}^T \bm{\mathcal{K}}^{-1} \bm{K} \right) } \,,
\end{eqnarray}
in Eq.~(\ref{eq:Mderiv}) we obtain
\renewcommand{\jot}{0.5em}
\begin{eqnarray} 
\lefteqn{ {}_n  \mathcal{M}_{i_1\, i_2 \, \cdots \, i_n} = 
\sqrt{\frac{R}{\delta \kappa (D)^2 \pi}}
 \, } \\
&& \hspace{6ex} \times \, \left( \frac{-1}{2 \bar{r}'}  \frac{\partial}{\partial V_{i_1}} \right) \left( \frac{-1}{2 \bar{r}'}  \frac{\partial}{\partial V_{i_2}} \right)
 \cdots \left( \frac{-1}{2 \bar{r}'}  \frac{\partial}{\partial V_{i_n}} \right) \,\exp{\left(- (K_0 - \bm{K}^T \bm{\mathcal{K}}^{-1}
\bm{K} ) \, \bar{r}^{\prime \, 2} \right) } \,,
 \nonumber \\
\end{eqnarray}
\renewcommand{\jot}{0em}
which yields
\begin{eqnarray}
{}_n  \mathcal{M}_{i_1\, i_2 \, \cdots \, i_n} & = & \sqrt{\frac{R}{\delta \kappa (D)^2 \pi}} \,\,\, \hat{C} \, \left( 
  {}_1\chir_{i_1} \, {}_1\chir_{i_2} \times \cdots \times {}_1\chir_{i_n}  \, \bar{r}^{\prime \, n} +
 {}_2\chir_{i_1 \,  i_2 } \, {}_1\chir_{i_3} \, {}_1\chir_{i_4} \times \cdots \times {}_1\chir_{i_n}  \, \bar{r}^{\prime \, n-2}  
\right. \nonumber\\&&
  + {}_2\chir_{i_1 \,  i_2 } \, {}_2\chir_{i_3 \, i_4 } \, {}_1\chir_{i_5} \, {}_1\chir_{i_6} \times \cdots \times {}_1\chir_{i_n}  \, \bar{r}^{\prime \, n-4} 
   \label{eqcalMn}
   \\
&&+ \cdots + \begin{array}{l@{}}
\left.{}_2\chir_{i_1 \,  i_2 } \, {}_2\chir_{i_3 \, i_4 }  \times \cdots \times {}_2\chir_{i_{n-1} \, i_n } \,\right) \mbox{\hspace{2ex} when $n$ is even} \,, \\
\left.{}_2\chir_{i_1 \,  i_2 } \, {}_2\chir_{i_3 \, i_4 }  \times \cdots \times  {}_2\chir_{i_{n-2} \, i_{n-1}} \,  {}_1\chir_{i_n} \bar{r}' \,\right)
\mbox{\hspace{2ex} when $n$ is odd} \,,
\end{array}  
\nonumber
\end{eqnarray}
where the $\chir s$ are elements of
\renewcommand{\jot}{0.5em}
\begin{eqnarray}
{}_1\bchir  & = &  - \frac{1}{2} \, \bm{\nabla}_{\bm{V}}  \, ( - R )
= \left( \begin{array}{c} 1 \\ - \bm{\mathcal{K}}^{-1} \bm{K} \end{array} \right) \\
{}_2\bchir & = & - \frac{1}{2} \, \bm{\nabla}_{\bm{V}} \otimes {}_1\bchir = \frac{1}{2} \left( \begin{array}{cc}
0 & 0 \\ 0 & \bm{\mathcal{K}}^{-1} \end{array} \right) \,,
\end{eqnarray}
\renewcommand{\jot}{0em}
${}_1\bchir$ is four-dimensional vector and ${}_2\bchir$ is a $4 \times 4$-dimensional matrix.
The $\hat{C}$ operator acts on each term in Eq.~(\ref{eqcalMn}) to produce a sum of terms over all
the distinct combinations of indices.

From Eq.~(\ref{eqcalMn}) we can now evaluate each element of the $\mathcal{M}$ tensors:
\renewcommand{\jot}{0.5em}
\begin{eqnarray}
{}_1  \mathcal{M}_{i} & = & \sqrt{\frac{R}{\delta \kappa (D)^2 \pi}} \,\,\, {}_1\chir_{i} \, \bar{r}'   \label{eq:1Mf}   \\
{}_3  \mathcal{M}_{i_1\, i_2 \,  i_3} & = & \, \sqrt{\frac{R}{\delta \kappa (D)^2 \pi}} \,\, \left( \, {}_1\chir_{i_1} \, {}_1\chir_{i_2} \,  {}_1\chir_{i_3}  \, \bar{r}^{\prime \, 3} +
( \,  {}_2\chir_{i_1 \,  i_2 } \, {}_1\chir_{i_3} + {}_2\chir_{i_1 \,  i_3 } \, {}_1\chir_{i_2}  + {}_2\chir_{i_2 \,  i_3 } \, {}_1\chir_{i_1}  \, )\, \bar{r}^{\prime}  \right) \,. \nonumber \\    \label{eq:3Mf} 
\end{eqnarray}
\renewcommand{\jot}{0em}
\subsection{Result: first-order density profile}
\subsubsection{density profile in displacement coordinates}
Using Eqs.~(\ref{eq:1Mf}) and (\ref{eq:3Mf}) in Eq.~(\ref{eq:N1SigmaCalM}) yields the density profile
\begin{equation} \label{eq:N1rA1A3}
  \mathcal{N}_1 (r) = \frac{N}{S(D)} \sqrt{\frac{R}{\delta \kappa (D)^2 \pi}} \, (1 + 
  \delta^{\frac{1}{2}} (A_1 \,  \bar{r}' + A_3 \, \bar{r}^{\prime \, 3}))) \exp (- R \bar{r}^{\prime \, 2}) \,,
\end{equation}
where
\begin{equation}
  \bar{r}' (r ; D) = \sqrt{D} \, \left( \frac{r}{\kappa (D)} -
  \bar{r}_{\infty} \right) \,.
\end{equation}

The coefficients $A_1$ and $A_3$ of the polynomial are
\renewcommand{\jot}{0.5em}
\begin{eqnarray}\label{eq:A1A3}
\label{eq:A1}  
A_1 &=&2 \, \left(\sum_i\,{}_{1}\chir_i\,{}_1\Xi_i + \sum_{i_1,i_2,i_3}\left( \,  {}_2\chir_{i_1 \,  i_2 } \, {}_1\chir_{i_3} + {}_2\chir_{i_1 \,  i_3 } \, {}_1\chir_{i_2}  + {}_2\chir_{i_2 \,  i_3 } \, {}_1\chir_{i_1} \right) \,{}_3\Xi_{i_1 \, i_2 \, i_3 } \right)
\\
\label{eq:A3}
  A_3 &=&2 \sum_{i_1,i_2,i_3}\,  {}_{1}\chir_{i_1} \, {}_{1}\chir_{i_2} \, {}_{1}\chir_{i_3
} \,{}_3\Xi_{i_1 \, i_2 \, i_3 } \,,
\end{eqnarray}
\renewcommand{\jot}{0em}
where
\renewcommand{\jot}{0.5em}
\begin{eqnarray}
  {}_1\Xi_{i} &=& \sum_{\mu=\{\mathbf{0\pm,1\pm}\}}T_{\mu,i}\,{}^1 E_{\mu}
\\
 {}_3\Xi_{i,j,k} &=&\sum_{\mu_1=\{\mathbf{0\pm,1\pm}\}}\sum_{\mu_2=\{\mathbf{0\pm,1\pm}\}}\sum_{\mu_3=\{\mathbf{0\pm,1\pm}\}} T_{\mu_1,i}T_{\mu_2,j}T_{\mu_3,k}
{}^3 E_{\mu_1, \mu_2, \mu_3}\,,
\end{eqnarray}
\renewcommand{\jot}{0em}
\begin{equation}\label{eq:X1}
{}_{1}\chir_i=
\left\{\begin{array}{ll}
       1 & i=1\\
	-(\mathcal{K}^{- 1} \mathbf{K})_{i-1} & 1 < i \leq 4
      \end{array}
\right.\,,
\end{equation}
and 
\begin{eqnarray}\label{eq:X3}
{}_{2}\chir_{i,j} \, {}_{1}\chir_{k}&=&
	\mathcal{K}_{i-1,j-1}^{- 1}\,{}_{1}\chir_k\Theta_{i-1}\Theta_{j-1}
+\mathcal{K}_{j-1,k-1}^{- 1}\,{}_{1}\chir_i\Theta_{j-1}\Theta_{k-1}
\nonumber\\&& 
+\mathcal{K}_{k-1,i-1}^{- 1}\,{}_{1}\chir_j\Theta_{k-1}\Theta_{i-1} \,,
\end{eqnarray}
where
\begin{equation}
\begin{array}{@{}rcll@{}}
\Theta_m & = & 0 & \mbox{when } m \leq 0 \\
 & = & 1 & \mbox{when } m \geq 1 \,.
\end{array}
\end{equation}
The first-order density profile has the form of a cubic polynomial multiplied by the lowest-order density profile. Note that the density profile is a function of the coordinate $r$, which is not the dimensionally-scaled internal displacement coordinate $\bar{r}' (r ; D)$. Thus one must make the following substitution to obtain the density profile as an explicit function of $r$:
\begin{equation}
  \bar{r}' (r ; D) = \delta^{- \frac{1}{2}} \left( \frac{r}{\kappa (D)} -
  \bar{r}_{\infty}\right)\,.
\end{equation}

We have derived the $N$-body density profile through first-order in $\delta^{1/2}$\,.
This density profile, $\mathcal{N}_1 \left(r \right)$\,, of Eq.~(\ref{eq:N1rA1A3})
includes the full interactions of $N$ particles through
first order in the perturbation series, exactly and analytically. The details of the interactions at each order are
folded into $R$\,, $A_1$\,, $A_3$\,, and $\bar{r}_{\infty}$\,.
We note though, that we have neglected the next-order term (order $\delta$) in Eq.~(\ref{eq:N1throughDelta2}).

\subsubsection{Density profile in oscillator units}
For the case of a system under harmonic confinement (such as the present trapped Hooke's law gas or a BEC in a parabolic trap) we may choose a oscillator-unit scaling $\kappa (D)=D^2 \bar{a}_{ho}$, where
\begin{equation}
\bar{a}_{ho}=\frac{1}{D^{3/2}}a_{ho} \,.
\end{equation}
Therefore
\begin{equation}
\kappa (D)=\sqrt{D}\,a_{ho}
\end{equation}
\begin{equation}
  \bar{r}'= ( \frac{r}{a_{ho}} - \sqrt{D}\,\bar{r}_{\infty})\,,
\end{equation}
and defining
\renewcommand{\jot}{0.5em}
\begin{equation}
 r_{osc}=\frac{r}{a_{ho}}
\end{equation}
\renewcommand{\jot}{0em}
in oscillator units, we obtain the Jacobian-weighted density per particle
\begin{eqnarray}
  \mathcal{N}_0 \left(r_{osc}\right)\times\frac{S(D)}{N}&=&\sqrt{\frac{R}{\pi}}
  \exp\left(- R\,\left(r_{osc}-\sqrt{D}\,\bar{r}_{\infty}\right)^2\right) \label{eq:N0result}
  \\
  \mathcal{N}_1 \left(r_{osc}\right)\times\frac{S(D)}{N} &=& 
  \left(1 + \delta^{\frac{1}{2}} \left(A_1  \left(r_{osc}-\sqrt{D}\,\bar{r}_{\infty}\right) + A_3 \left(r_{osc}-\sqrt{D}\,\bar{r}_{\infty}\right)^3\right)\right)
\nonumber  \\&&\times \mathcal{N}_0 \left(r_{osc}\right)\times\frac{S(D)}{N} \,. \label{eq:N1result}
\end{eqnarray}
\renewcommand{\jot}{0em}

\section{Test Application: Harmonically Interacting Particles Under Harmonic Confinement}
\label{sec:test}
In this paper we have derived the Jacobian-weighted density profile through first order from the wave function through first
order as an example of an observable which may be derived from the $N$-body, interacting wave function.
We test this general formalism for the density profile, by comparing it to the density profile
of the analytically solvable system of $N$\,, harmonically-interacting particles in a harmonic confining potential with Hamiltonian

\begin{equation}
H = \frac{1}{2} \, \left( \sum_i^N  \left[ - \frac{\partial^2}{\partial \bm{r}_i^2} + \omega_t^2 \bm{r}_i^2 \right]  + \sum_{i<j} \omega_p^2 \bm{r}_{i,j}^2 \right) \,.
\end{equation}
The exact analytic density profile for this system is is independently derived in Appendix~\ref{app:dens} (see Eqs.~(\ref{eq:Nr}) and (\ref{eq:Nbarr})), and this is expanded
through first order in $\delta^{1/2}$ to yield the exact density profile through first order
(see Eq.~(\ref{eq:Ndensharm1})). This analysis shows that the density profile for any $N$ or interaction
strength follows a universal curve when a simple scaling is applied to the radial variable (it should be noted
that this is not true of the wave function) and is given by
\begin{equation} \label{eq:NbarreffMain}
\mathcal{N}(\bar{r}_{\rm eff}) = \frac{2 \, D^{\frac{D}{2}} } {\Gamma \hspace{-0.4ex} \left( \frac{D}{2} \right) }
\,\, \bar{r}_{\rm eff}^{D-1} \, \exp{ \left(  - D \,\,\bar{r}^2_{\rm eff} \right)   } \,,
\end{equation}
where
\begin{equation}
\bar{r}_{\rm eff} = \sqrt{\lambda_{\rm eff}} \, \bar{r} \,,
\end{equation}
\begin{equation}
\lambda_\textrm{eff} =  \frac{N \lambda}{N+\lambda -1}\,, \mbox{\hspace{3ex}}
\lambda = \sqrt{1+N \lambda_p^2} \,, \mbox{\hspace{3ex}} \lambda_p = \omega_p / \omega_t
\end{equation}
and
\begin{equation}
\mathcal{N}(\bar{r}'_{\rm eff}) = \frac{1}{\sqrt{\lambda_{\rm eff}}} \, \mathcal{N}(\bar{r}') \,.
\end{equation}
Expanded to first order Eq.~(\ref{eq:NbarreffMain}) gives
\begin{equation} \label{eq:Ndensharm1Main}
\mathcal{N}_1(\bar{r}'_{\rm eff}) = \left(  1 + \delta^{\frac{1}{2}} \, \sqrt{2} \, \left(  \frac{2 \, \bar{r}_{\rm eff}^{\prime 3} }{3 } 
- \bar{r}'_{\rm eff}  \right)  \right)
\left( \frac{2 } {\pi} \right)^{\frac{1}{2}} \, \exp{ \left(  -2 \, \bar{r}_{\rm eff}^{\prime 2}   \right)  } \,.
\end{equation}

\begin{figure}[p]
\includegraphics[scale=1]{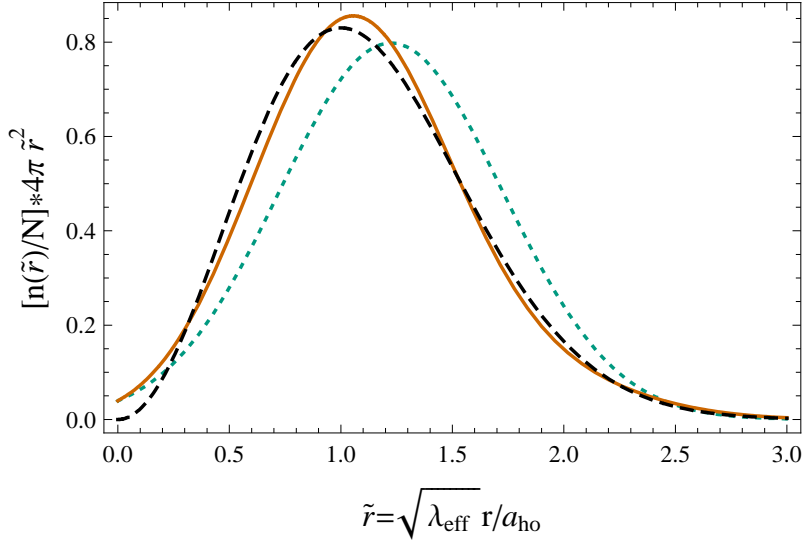}
\caption{Scaled density profile at $D=3$ for $N$ harmonically-interacting particles under harmonic confinement in oscillator units of the confining potential. The short dash curve is the lowest-order density profile, while the solid curve is the density profile through first order. The long-dash curve is the exact result. The scaling factor, $\sqrt{\lambda_{\rm eff}}$ is explained in Appendix~\ref{app:dens}. }
\label{fig:scaled}
\end{figure}

This scaled density profile for $D=3$ ($\delta = 1/3$) is plotted in Fig.~\ref{fig:scaled}.
One readily sees the improvement obtained at first order, confirming the efficacy of this
approach to the general confined $N$-body problem, which may be systematically improved by going to
higher orders.

\begin{figure}[p]\centering
\includegraphics[scale=1]{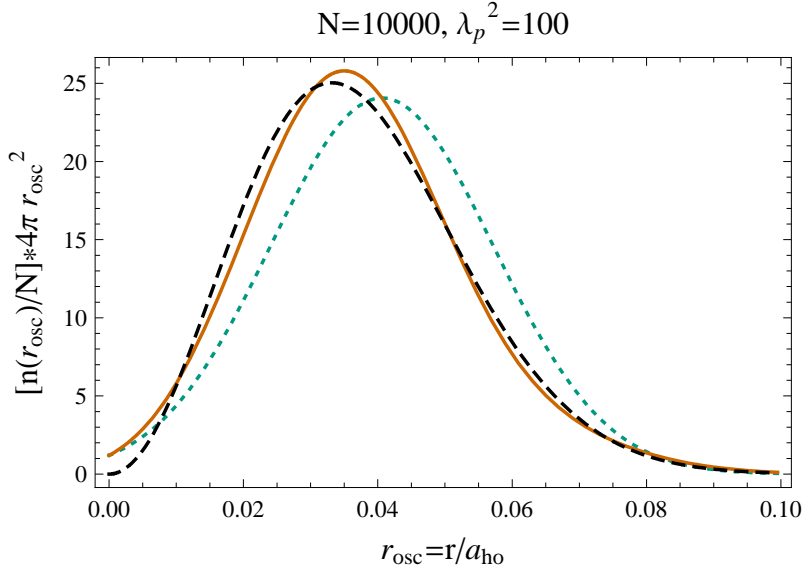}
\caption{Unscaled density profile at $D=3$ for $N=10,000$ particles under harmonic confinement with strong attractive
harmonic interactions ($\lambda_p^2 = 100$) in oscillator units of the confining potential. The short dash curve is the lowest-order density profile, while the solid curve is the density profile through first order. The long-dash curve is the exact result. The parameter $\lambda_p^2$\,, as explained in Appendix~\ref{app:dens}, is the interaction frequency squared in oscillator units of the confining potential. }
\label{fig:att}
\end{figure}

\begin{figure}[p]\centering
\includegraphics[scale=1]{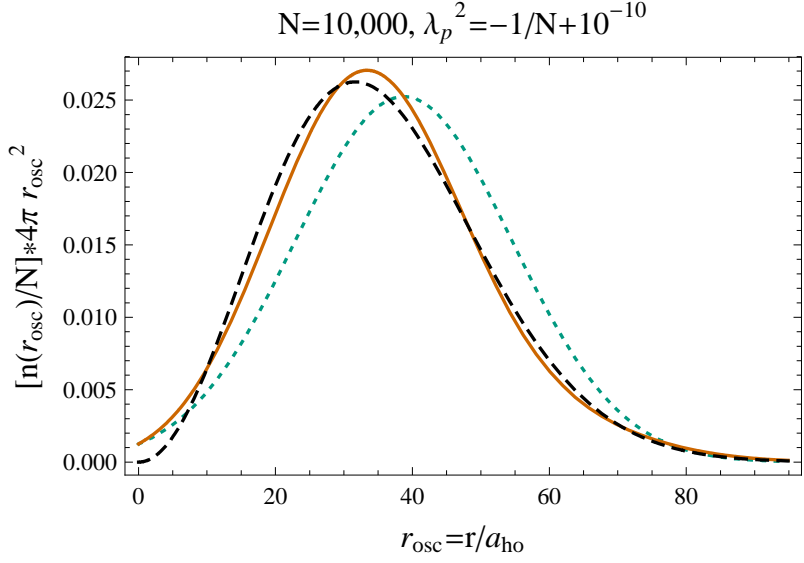}
\caption{Unscaled density profile at $D=3$ for $N=10,000$ particles under harmonic confinement with repulsive
harmonic interactions ($\lambda_p^2 = -1/10,000 + 10^{-10}$) in oscillator units of the confining potential. The system is just below the dissociation threshold and very extended. The short dash curve is the lowest-order density profile, while the solid curve is the density profile through first order. The long-dash curve is the exact result. The parameter $\lambda_p^2$\,, as explained in Appendix~\ref{app:dens}, is the interaction frequency squared in oscillator units of the confining potential. }
\label{fig:rep}
\end{figure}

The general theory developed in this
paper for the density profile involves no such harmonic interaction specific scaling since it's applicable
to any interparticle potential, not just harmonic
interparticle potentials. Consequently in Figs.~\ref{fig:att} and \ref{fig:rep} we plot the density
profile for $D=3$ ($\delta = 1/3$) without this harmonic-interaction specific scaling for two very different interparticle interaction strengths. Both are for
$N=10,000$ particles, but Fig.~\ref{fig:att} features strongly attractive interactions, while Fig.~\ref{fig:rep}
is for a repulsive interaction just below the dissociation limit.
In the former case the system is tightly bound and very compact. In the latter case the confining potential
is barely able to hold the system together against the combined effect of the repulsive interactions,
and the system is very extended.

The density profile derived from the general $N$-body formalism developed in this paper, and implemented
in \emph{Mathematica}\cite{mathematica} code\cite{MATprog}, is indistinguishable from the density profiles
derived from the exact independent solution of the harmonically-interacting system. The agreement between the general formalism,
which uses a perturbation series invariant under $S_N$\,, and the direct density profiles of Eqs.~(\ref{eq:NbarreffMain}) and
(\ref{eq:Ndensharm1Main}) obtained in Appendix~\ref{app:dens}, confirms the correctness of the general formalism developed in this paper, and its implementation in
\emph{Mathematica}\cite{mathematica} code.

\section{Summary and Conclusions}
\label{sec:Concl}
While the resources required to solve classical systems scale as a polynomial with the number of particles, $N$\,, allowing calculations to be performed involving large numbers of particles, the situation regarding large-$N$ quantum systems is not so felicitous. In this case the resources required scale exponentially with $N$\,, making calculations for large-$N$ quantum systems a far more formidable challenge, unless $N$ and the
interparticle interaction strengths allow an approximation, such as the mean-field approximation.

In a series of papers, we have been developing an approach to the general interacting quantum $N$-body problem, which while essentially analytic, makes no assumtions regarding the form or strength of the interparticle interactions. This approach derives the interacting $N$-body wave function,
from which any observable quantity can in principle be derived. In the process,
collective, normal mode coordinates are derived revealing the nature
of the microscopic motions of the particles of the system for any interaction. Using this wave function, properties such as energies and density profiles have been derived.

The method involves expanding the system in inverse powers of the spatial dimension $D$\,. At large $D$\,, systems exhibit a point
group structure of a far higher degree of symmetry (isomorphic to $S_N$) than is possible in three dimensions, allowing group theory and graphical techniques to be used to tame the exponential $N$ scaling, leading to an essentially analytic solution at lowest order in the wave function.

More recently, in a major development of this approach, this method has been succesfully extended, analytically and exactly, to first-order in the wave function, and in principle the techniques developed to do this can be extended to yet higher orders.

This paper is the first application of this first-order wave function to the derivation of physical property of interacting $N$-body quantum
systems, namely the density profile to first order. In a test of this theory, the derived first-order density profile is tested on an exactly
solvable model, namely a system of harmonically interacting particles in a harmonic confining potential. The harmonic interparticle interactions
may be attractive, or repulsive, and if the interactions are sufficiently repulsive the system will dissociate despite the presence of the harmonic
confining potential.

The general theory developed in this and prior papers, agrees with the exact results for this system obtained from the independent solution,
showing strong convergence to the exact, {\em three-dimensional} result for both strongly attractive interactions and repulsive interactions just below the dissociation threshold.

While this paper (as well as Ref.~\onlinecite{wavefunction1harm}) has focused on the harmonically-interacting system in a harmonic confining
potential, the theory is not limited to these systems and in past papers we have examined other systems at lowest order, such as the
Bose-Einsten condensate, and quantum dots. While the lowest-order approximation for the BEC captures the behavior of the system for a range
of $N$ and interaction strengths, when $N$ or the interaction strength is larger than this range, the lowest-order density profile increasingly
does not have the flexiblity needed to capture the behavior of the system. The first-order result does not have this limitation and so it is
very desirable to apply the first-order density profile derived in this paper to strongly interacting BECs as well as other strongly interacting systems. Of particular note is the fact that the functional form of the first-order density profile of Eq.~(\ref{eq:N1result}) admits structure (wiggles) indicating the onset of crystallization/fermionization. Although such behavior is not seen for the long-range harmonic interactions
examined in this paper, other systems do exhibit such transitions.

While we focused in this paper on density profiles, the theory derives the exact, first-order wave function from which any observable may be derived.
Also excited states may be derived, and the theory needed to extend these results to higher-angular-momentum states has been set
up\cite{highLI,highLII}.

\begin{acknowledgments}
We gratefully acknowledge continued support from the Army Research Office.
\end{acknowledgments}

\newpage

\appendix

\section{Binary Invariants}
\label{app:Bin}
In Ref.~\onlinecite{wavefunction1} we introduce binary invariants which are binary tensors invariant under the $S_N$ particle
interchange group.
\subsection{Introducing Graphs}
\begin{definition}
A \emph{graph} $\mathcal{G} =(V,E)$ is a set of vertices $V$ and
edges $E$. Each edge has one or two associated vertices, which are
called its \emph{endpoints}.
\end{definition}

For example, \graphggb is a graph $\mathcal{G}$ with three vertices (or "dots'') and two
edges (or lines). We allow our graphs to include loops and multiple edges\cite{multigraph}.
A graph contains information regarding the connectivity of edges and vertices only: the
orientation of edges and vertices is insignificant. 

We introduce a mapping which associates each tensor element with a graph as
follows:
\begin{enumerate}
\item draw a labeled vertex ($\labeledgraphr{i}$) for each distinct index in the set of indices of the element
\item draw an edge ($\labeledgraphgamma{i}{j}$) for each double index $(ij)$
\item draw a ``loop'' edge ($\labeledgraphr{i}$) for each distinct single index $i$
\end{enumerate} 
For example, the graph corresponding to the tensor element $\stacklr{0}{2}{Q}{r\gamma}{i,(ij)}$ under this mapping is $\labeledgraphgra{i}{j}$\,.

Two graphs with the same number of vertices and edges that are connected the same way are called \emph{isomorphic}. The action $S_N$ group interchanging particle labels, only connects tensor elements
with isomorphic graphs; two elements with heteromorphic graphs are never connected by the $S_N$
group. Now consider a tensor, for which all of the elements labeled by a a single isomorphic set of
graphs are equal to unity, while all of the other elements labeled by graphs heteromorphic to this single
set of isomorphic graphs are equal to zero. We we term this tensor a binary invariant,
$[B ({\mathcal G})]_{\nu_1, \nu_2, \ldots}$ since it
is invariant under the $S_N$ group, and we label it by the graph $\mathcal{G}$\., sans particle labels
at the vertices, for the non-zero elements all of which are equal to unity.

Denoting the set of unlabeled graphs for each block as $\mathbb{G}_{X_1X_2\ldots X_n}$, where $n$ is the rank of the tensor block (and therefore the number of edges in each graph in the set) and $X$ is $r$ or $\gamma$\,,
we have
\begin{eqnarray}\label{eq:GXX}
\mathbb{G}_{rr}&=&\{\graphrra,\graphrrb\}
\nonumber\\
\mathbb{G}_{\gamma r}&=&\{\graphgra,\graphgrb\}
\\
\mathbb{G}_{\gamma \gamma}
&=&\{\graphgga,\graphggb,\graphggc\}
\nonumber
\end{eqnarray}
\begin{eqnarray}\label{eq:GXXX}
\mathbb{G}_{r}&=&\{\graphr\}
\nonumber\\
\mathbb{G}_{\gamma}&=&\{\graphgamma\}
\nonumber\\
\mathbb{G}_{rrr}&=&\{\graphrrra,\graphrrrb,\graphrrrc\}
\\
\mathbb{G}_{\gamma rr}
&=&\{\graphgrra,\graphgrrb,\graphgrrc,\graphgrrd,\graphgrre\}
\nonumber\\
\mathbb{G}_{\gamma \gamma r}
&=&\{\graphggra,\graphggrb,\graphggrc,\graphggrd,\graphggre,\graphggrf,\graphggrg\}
\nonumber\\
\mathbb{G}_{\gamma \gamma \gamma}
&=&\{\graphggga,\graphgggb,\graphgggc,\graphgggd,\graphggge,\graphgggf,\graphgggg,\graphgggh\}
\nonumber
\end{eqnarray}
Each of these graphs denotes a binary invariant.

\section{Density Profile of Harmonically Interacting Particles Subject to a Harmonic Confining Potential: An Independent Solution}
\label{app:dens}
The Hamiltonian of the harmonically-interacting model system of identical particles is
\begin{equation}
H = \frac{1}{2} \, \left( \sum_i^N  \left[ - \frac{\partial^2}{\partial \bm{r}_i^2} + \omega_t^2 \bm{r}_i^2 \right]  + \sum_{i<j} \omega_p^2 \bm{r}_{i,j}^2 \right) \,.
\end{equation}

\subsection{The exact $N$-body Wave function}
Making the orthogonal transformation to center-of-mass and Jacobi coordinates
\begin{eqnarray}
\bm{R} = \frac{1}{\sqrt{N}} \sum_{k=1}^N \bm{r}_k & \hspace{1ex} \mbox{and} \hspace{1ex} & \bm{\rho}_i = \frac{1}{\sqrt{i(i+1)}} \left( \sum_{j=1}^i \bm{r}_j- i \bm{r}_{i+1} \right) \,,
\end{eqnarray}
where $1 \leq i \leq N-1$\,, the Hamiltonian becomes
\begin{equation}
H = \frac{1}{2} \, \left( - \frac{\partial^2}{\partial \bm{R}^2} + \omega_t^2 \bm{R}^2 \right)
+ \frac{1}{2} \, \sum_{i=1}^{N-1} \left( - \frac{\partial^2}{\partial \bm{\rho }_i^2} + \omega_{\rm int}^2 \bm{\rho}_i^2 \right) \,,
\end{equation}
the sum of $N$\,, $D$-dimensional, harmonic-oscillator Hamiltonians, where
\begin{equation}
\omega_{\rm int} = \sqrt{\omega_t^2 + N\, \omega_p^2}\,.
\end{equation}
The ground-state solution of the Schr\"odinger equation
\begin{equation}
 H \, \Psi = E \, \Psi
\end{equation}
is the product of harmonic-oscillator wave functions
\begin{equation}
\label{eq:WFNJ}
\Psi(\bm{R}, \, \{\bm{\rho}_i\}; \, D) = \psi(R;\,  \omega_t, \, D) \, \prod_{i=1}^{N-1} \psi(\rho_i; \, \omega_{\rm int }, \, D) \,,
\end{equation}
where $\psi(r; \, \omega, \, D)$ is the $D$-dimensional, harmonic-oscillator, ground-state wave function
\begin{equation}
\label{eq:DHarmWF}
\psi(r; \, \omega, \, D) = \sqrt{\frac{ \omega^{\frac{D}{2}}}{\pi^{\frac{D}{2}} }} \,\, \exp{\left( - \frac{\omega}{2} r^2 \right) }
\end{equation}
and $\psi(r; \, \omega, \, D)$ satisfies the normalization condition
\begin{equation}
\int_0^\infty [\psi(r; \, \omega, \, D)]^2 \,r^{D-1}\,d^D\bm{r} = 1 
\end{equation}
so that
\begin{equation}
\int_{-\infty}^\infty [\Psi(\bm{R}, \, \{\bm{\rho}_i\}; \, D)]^2 \,  \prod_{i=1}^{N-1} d^D\bm{\rho}_i \, d^D\bm{R} = 1 \,.
\end{equation}

\subsection{The exact $N$-body density profile}
Since the wave function is a completely symmetric function under any permutation of the particles, one can
write the Jacobian-weighted density profile as
\begin{equation}
\mathcal{N}(r) = \int_{-\infty}^\infty [\Psi(\bm{R}, \, \{\bm{\rho}_i\}; \, D)]^2 \, \delta(r-r_N) \, \,  \prod_{i=1}^{N-1} d^D\bm{\rho}_i \, d^D\bm{R} \,,
\end{equation}
where from Eqs~(\ref{eq:WFNJ}) and (\ref{eq:DHarmWF}), the unweighted wave function is
\begin{equation}
\Psi(\bm{R}, \, \{\bm{\rho}_i\}; \, D) = \sqrt{\frac{ \omega_t^{\frac{D}{2}}}{\pi^{\frac{D}{2}} }} \,\, \exp{\left( - \frac{\omega_t}{2} \bm{R}^2 \right) } \,\,  \left( \frac{ \omega_{\rm int }^{\frac{D}{2}}}{\pi^{\frac{D}{2}} } \right)^{\frac{N-1}{2}} \prod_{i=1}^{N-1}  \,\, \exp{\left( - \frac{\omega_{\rm int }}{2} \bm{\rho}_i^2 \right) } \,.
\end{equation}
Since the transformation from single-particle coordinates, $\bm{r}_1$\,, \ldots, $\bm{r}_N$\,, to
Jacobi/center-of-mass coordinates, $\{\bm{\rho}_i\} / \bm{R}$\,, is orthogonal,
\begin{equation} \label{eq:NM}
\mathcal{N}(r) = \left( \frac{ \omega_t^{\frac{D}{2}}}{\pi^{\frac{D}{2}} } \right) \, 
\left( \frac{ \omega_{\rm int }^{\frac{D}{2}}}{\pi^{\frac{D}{2}} } \right)^{N-1}
 M(r)  = r^{D-1} \, \rho(r) 
\end{equation}
where $\rho(r)$ is the unweighted density profile, and
\begin{equation}
M(r) = \int d^D\bm{r}_1 \, d^D\bm{r}_2 \, \ldots \, d^D\bm{r}_N \, \delta(r-r_N) \,
\exp{\left( - (\omega_t - \omega_{\rm int }) \bm{R}^2 \right) } \, 
\exp{\left( - \omega_{\rm int } \left( \bm{R}^2 + \sum_{i=1}^{N-1} \bm{\rho}_i^2 \right) \right) } \,.
\end{equation}
Defining
\begin{equation}
_{N}\bm{r} = \bm{r}_1 \oplus \bm{r}_2 \oplus \cdots \oplus \bm{r}_N =\left(\begin{array}{c}
\bm{r}_1\\\bm{r}_2\\\vdots\\\bm{r}_N \end{array}\right)
\,,
\end{equation}
\begin{equation}
_{N}\bm{J} = \mathbb{J}_N \otimes \bm{I}_D \,,
\end{equation}
where $\bm{I}_D$ is the $D \times D$ dimensional unit matrix in the $D$-dimensional coordinate space
and $\mathbb{J}_N$ is the $N \times N$ dimensional matrix of elements equal to one in the particle label space,
then we can write
\begin{equation}
\bm{R}^2 = \frac{1}{N} \,  { _{N}\bm{r}^T}  {}_N\bm{J} \, _{N}\bm{r}
\end{equation}
\begin{equation}
\bm{R}^2 + \sum_{i=1}^{N-1} \bm{\rho}_i^2 = \sum_{i=j}^N \bm{r}_j^2 = \, { _{N}\bm{r}^T}  _N\bm{I} \, _{N}\bm{r} \,,
\end{equation}
where
\begin{equation}
_N\bm{I} = \mathbb{I}_N \otimes \bm{I}_D \,,
\end{equation}
and $\mathbb{I}_N$ is the $N \times N$ dimensional unit matrix in the particle label space.
Since $\int d^D\bm{r}_N = \int r_N^{D-1} dr_N \int d\bm{\Omega}_N$, where $\int d\bm{\Omega}_N$ is the integral over
the $D$-dimensional solid angle, we can write
\begin{eqnarray}
M(r) & = & \int d^D\bm{r}_1 \, d^D\bm{r}_2 \, \ldots \, d^D\bm{r}_N \, \delta(r-r_N) \,
\exp{\left( -  \frac{(\omega_t - \omega_{\rm int })}{N} \,  { _{N}\bm{r}^T}  {}_N\bm{J} \, _{N}\bm{r} \right) } \, 
\exp{\left( - \omega_{\rm int } \,\,  { _{N}\bm{r}^T}  _N\bm{I} \, _{N}\bm{r} \right) }  \nonumber  \\
& = & S(D) \, r^{D-1} \, \exp{\left( -  \frac{(\omega_t + (N-1) \omega_{\rm int })}{N} \, \bm{r}^2 \right) } \nonumber  \\  
&&\times \int d^D\bm{r}_1 \, d^D\bm{r}_2 \, \ldots \, d^D\bm{r}_{N-1} 
\exp{\left( -   { _{N-1}\bm{r}^T}  \bm{A} \,\, _{N-1}\bm{r} - 2 \, \bm{B}^T \, _{N-1}\bm{r} \right) } \,, \label{eq:m_int}
\end{eqnarray}
where
\begin{equation} \label{eq:solid_ang}
S(D) = \frac{2 \, \pi^{\frac{D}{2}}}{\Gamma\left( \frac{D}{2}\right)}
\end{equation}
is the $D$-dimensional solid angle,
\begin{equation}
_{N-1}\bm{r} = \bm{r}_1 \oplus \bm{r}_2 \oplus \cdots \oplus \bm{r}_{N-2} \oplus \,\bm{r}_{N-1} \,,
\end{equation}
\renewcommand{\jot}{0.5em}
\begin{eqnarray}
\bm{A} & = &   \omega_{\rm int } \,\,    _{N-1}\bm{I}  \,\, + \,\,  \frac{(\omega_t - \omega_{\rm int })}{N} \,\,  \,_{N-1}\bm{J}  \\
\bm{B} & = &  \frac{(\omega_t - \omega_{\rm int })}{N} \,\,  _c\bm{1}  \,\, \bm{r} \,  \label{eq:B}
\end{eqnarray}
\renewcommand{\jot}{0em}
\begin{equation}
\,_{N-1}\bm{J} = \mathbb{J}_{N-1} \otimes \bm{I}_D  \,,
\end{equation}
$\mathbb{J}_{N-1}$ is the $(N-1) \times( N-1)$-dimensional matrix of elements equal to one in the particle label space of the first $N-1$ particles,
\begin{equation}
_{N-1}\bm{I} = \mathbb{I}_{N-1} \otimes \bm{I}_D  \,,
\end{equation}
and $\mathbb{I}_{N-1}$ is the $(N-1) \times( N-1)$-dimensional unit matrix in the particle label space of the first $N-1$ particles,
\begin{equation}
_c\bm{1} = \mathbf{1}_c \otimes \bm{I}_D  \,,
\end{equation}
where $\mathbf{1}_c$ is the $( N-1)$-dimensional column vector of elements equal to one in the particle label space of the first $N-1$ particles so that
\begin{equation}
_c\bm{1}^T \, _{N-1}\bm{r} \, = \,  \sum_{k=1}^{N-1} \bm{r}_k \,.
\end{equation}
Using the result
\begin{equation}
\int_{-\infty}^\infty d^n\bm{b} \exp{\left( -   { \bm{b}^T}  \bm{A} \, \bm{b} - 2 \, \bm{B}^T \, \bm{b} \right) }
=\frac {\pi^\frac{n}{2} }{\sqrt{\det{\bm{A}}} }  \, \exp{ \left( \bm{B}^T \bm{A}^{-1} \, \bm{B} \right)  }
\end{equation}
in Eq.~(\ref{eq:m_int}), with the identification $n=D(N-1)$ and $\bm{b}=\bm{r}$\,, we obtain
\begin{equation} \label{eq:Mint}
M(r) = S(D) \, r^{D-1} \, \exp{\left( -  \frac{(\omega_t + (N-1) \omega_{\rm int })}{N} \,  \bm{r}^2 \right) } \, \frac {\pi^\frac{D(N-1)}{2} }{\sqrt{\det{\bm{A}}} }  \, \exp{ \left( \bm{B}^T \bm{A}^{-1} \, \bm{B} \right)  } \,.
\end{equation}
Thus we need to evaluate $\det{\bm{A}}$ and $\bm{A}^{-1}$\,. First $\bm{A}^{-1}$\,. One has that
$(\bm{U} \otimes \bm{V})^{-1} = \bm{U}^{-1} \otimes \bm{V}^{-1}$\,. Thus writing
\begin{equation} \label{eq:V}
\bm{U}  =   \alpha \,\,  \mathbb{I}_{N-1}  \,\, + \,\, \beta \,\,    \, \mathbb{J}_{N-1} \,,
\end{equation}
where $\alpha = \omega_{\rm int }$ and $\beta = \frac{(\omega_t - \omega_{\rm int })}{N}$\,, and using the closed algebra
\begin{eqnarray}
\,\mathbb{J}_{N-1} \,\, \,\mathbb{J}_{N-1} & = & (N-1) \,\, \,\mathbb{J}_{N-1} \nonumber \\
\,\mathbb{J}_{N-1} \,\, \mathbb{I}_{N-1} & = & \mathbb{I}_{N-1} \,\, \,\mathbb{J}_{N-1}  =  {}\,\mathbb{J}_{N-1} \label{eq:alg} \\
\mathbb{I}_{N-1} \,\, \mathbb{I}_{N-1} & = & \mathbb{I}_{N-1} \,, \nonumber
\end{eqnarray}
we obtain
\begin{equation}
\bm{U}^{-1}  =   (\alpha \,\,  \mathbb{I}_{N-1}  \,\, + \,\, \beta \,\,    \,\mathbb{J}_{N-1} )^{-1}
= \frac{1}{\alpha} \left(  \mathbb{I}_{N-1} - \frac{\beta}{\alpha + (N-1) \beta} \,\, \,\mathbb{J}_{N-1}  \right) \,.
\end{equation}
So with $\bm{V} = \bm{I}_D$\,, one finds that
\begin{equation} \label{eq:Ai}
\bm{A}^{-1} = \frac{1}{\omega_{\rm int }} \left(  \mathbb{I}_{N-1} \,\, -  \,\,  \frac{ (\omega_t - \omega_{\rm int })}{ (N-1) \, \omega_t + \omega_{\rm int }  }  \,\, \,\mathbb{J}_{N-1}  \right) \otimes \bm{I}_D  \,.
\end{equation}
To evaluate $\det{\bm{A}}$ we note that
\begin{equation} \label{eq:detUoV}
\det{(\bm{U} \otimes \bm{V})} = (\det{\bm{U}})^{d_V}  \, (\det{\bm{V}})^{d_U} \,,
\end{equation}
where $d_U$ and $d_V$ are the dimensionalities of matrices $\bm{U}$ and $\bm{V}$ respectively. The determinant of $\bm{I}_D$ is simple enough and equals unity. For $\det{\bm{U}}$ of Eq.~(\ref{eq:V}), we show that
\begin{equation} \label{eq:detV}
\det{\bm{U}} = \det{ \left[   \alpha \,\,  \mathbb{I}_{N-1}  \,\, + \,\, \beta \,\,    \,\mathbb{J}_{N-1}  \right]   }
= \alpha^{(N-2)} \, (\alpha + (N-1) \, \beta )
\end{equation}
\subsubsection*{Proof:}
We have
\begin{equation}
\det{ \left[   \alpha \,\,  \mathbb{I}_{N-1}  \,\, + \,\, \beta \,\,    \,\mathbb{J}_{N-1}  \right]   } = \| \bm{E} \| \,,
\end{equation}
where $\bm{E}$ is the diagonal matrix of eigenvalues of $\bm{V}$, and $\| \bm{E} \|$ is the norm of $\bm{E}$\,,
the product of its eigenvalues. Now since $\mathbb{J}_{N-1} = \bm{1}_c \bm{1}_c^T$\,, $\bm{x}_{N-1} =\frac{1}{\sqrt{N-1}} \, \bm{1}_c$ is seen to be a normalized eigenvector of $\bm{U}$ satisfying
\begin{equation}
\bm{U} \, \bm{x}_{N-1} = E_{N-1} \, \bm{x}_{N-1} \,,
\end{equation}
where
\begin{equation}
E_{N-1} = \alpha + (N-1) \beta \,.
\end{equation}
The remaining $N-2$ eigenvectors, $\bm{x}_j$\,, $\forall \hspace{1em} 1 \leq j \leq N-2$\,, are orthogonal to
$\bm{x}_{N-1}$ so $\mathbb{J}_{N-1} \, \bm{x}_j = 0$ $\hspace{1em} \forall \hspace{1em} 1 \leq j \leq N-2$\,, from which it follows
that
\begin{equation}
\bm{U} \, \bm{x}_j = E_j \, \bm{x}_j \,,
\end{equation}
where
\begin{equation}
E_j = \alpha  \hspace{2em} \forall \hspace{1em} 1 \leq j \leq N-2 \,.
\end{equation}
Putting it all together we arrive at Eq.~(\ref{eq:detV}). Q.E.D..

Using Eq.~(\ref{eq:detV}) in Eq.~(\ref{eq:detUoV}) we arrive at
\renewcommand{\jot}{0.5em}
\begin{eqnarray} \label{eq:detA}
\det{\bm{A}} & = & \alpha^{D(N-2)} \, (\alpha + (N-1) \, \beta )^D \\
 & = & \frac{\omega_{\rm int }^{D(N-2)} } {N^D} \, ( (N-1) \omega_t +  \omega_{\rm int } )^D
\end{eqnarray}
\renewcommand{\jot}{0em}
As for $ \exp{ \left( \bm{B}^T \bm{A}^{-1} \, \bm{B} \right)  }$\,, Eqs.~(\ref{eq:B}) and (\ref{eq:Ai}), along with
\begin{eqnarray}
\bm{1}_c^T \, \mathbb{I}_{N-1} \, \bm{1}_c & = &  (N-1)  \\
\bm{1}_c^T \, \mathbb{J}_{N-1} \, \bm{1}_c & = &  (N-1)^2 \,,
\end{eqnarray}
give us
\begin{equation} \label{eq:expBTAiB}
\exp{ \left( \bm{B}^T \bm{A}^{-1} \, \bm{B} \right)  } = \exp{ \left(  \frac{ (N-1) \, (\omega_t - \omega_{\rm int })^2} { (N-1) \omega_t + \omega_{\rm int }  }   \, \bm{r}^2
\right)  } \,.
\end{equation}
Thus from Eqs.~(\ref{eq:Mint}), (\ref{eq:detA}), and (\ref{eq:expBTAiB}) we obtain
\begin{equation} \label{eq:Mf}
M(r) = S(D) \, r^{D-1} \, \frac {\pi^\frac{D(N-1)}{2} N^\frac{D}{2}}{\sqrt{\omega_{\rm int }^{D(N-2)}  \, ( (N-1) \omega_t +  \omega_{\rm int } )^D} } \, \exp{\left( -  \omega_{\rm eff } \,  \bm{r}^2 \right) }  \,,
\end{equation}
where
\begin{equation} \label{eq:weff}
\omega_{\rm eff } = \frac{N \omega_t \, \omega_{\rm int } } { (\omega_{\rm int } + (N-1) \, \omega_t)} \,.
\end{equation}
Finally using Eqs.~(\ref{eq:solid_ang}), (\ref{eq:Mf}),  and (\ref{eq:weff}) in Eq.~(\ref{eq:NM}) we obtain
\begin{equation} \label{eq:Nr}
\mathcal{N}(r) = \frac{2 \, (\lambda_{\rm eff}\omega_t)^{\frac{D}{2}} } {\Gamma \hspace{-0.4ex} \left( \frac{D}{2} \right) }
\,\, r^{D-1} \, \exp{ \left(  - \lambda_{\rm eff}\omega_t \, r^2 \right)   } \,,
\end{equation}
where
%
%
%
\begin{equation}
\lambda_{\rm eff} = \frac{N \omega_{\rm int} } { \omega_{\rm int} + (N-1) \omega_t } = \frac{1}{2 \bar{r}_\infty^2} \,.
\end{equation}
Note that
\begin{equation}
\int_0^\infty \mathcal{N}(r) dr = 1\,.
\end{equation}

\subsection{Dimensional Expansion of the Number Density}
\subsubsection{Dimensional scaling}
To obtain the dimensional expansion of the density profile of Eq.~(\ref{eq:Nr}), we first need to regularize the large-dimension limit by dimensionally scaling
the parameters and variables. As in Ref.~\onlinecite{GFpaper}, we define dimensionally scaled frequency $\bar{\omega}_t$ and the 
%
%
 dimensionally-scaled, oscillator-scaled radial variable $\bar{r}$,
\begin{eqnarray}
\bar{\omega}_t&=&D^3\omega_t\\
\bar{r} &=& \sqrt{\bar{\omega}_t} \, \frac{r}{D^2} = \sqrt{\frac{\omega_t}{D}} \, r \,,
\end{eqnarray}
from which we derive
%
%
%
\begin{equation}
(\lambda_{\rm eff}\omega_t)^{\frac{D}{2}} \, r^{D-1} dr = (\lambda_{\rm eff} \, D)^{\frac{D}{2}} \, \bar{r}^{D-1} d\bar{r} \,.
\end{equation}
Thus the number density in dimensionally-scaled coordinates is
\begin{equation} \label{eq:Nbarr}
\mathcal{N}(\bar{r}) = \frac{2 \, (\lambda_{\rm eff} \, D)^{\frac{D}{2}} } {\Gamma \hspace{-0.4ex} \left( \frac{D}{2} \right) }
\,\, \bar{r}^{D-1} \, \exp{ \left(  - \lambda_{\rm eff} \, D \,\,\bar{r}^2 \right)   } \,,
\end{equation}
where
\begin{equation} \label{eq:normbar}
\int_0^\infty \mathcal{N}(\bar{r}) d\bar{r} = 1\,.
\end{equation}

Equation~(\ref{eq:Nbarr}) implies (as does Eq.~(\ref{eq:Nr})) that up to a scaling, the density profiles for
harmonically-interacting particles in a harmonic-confining potential follow a universal curve for any
$N$ or interparticle interaction strength in oscillator units $\lambda_p \equiv \omega_p / \omega_t$\,, where
\begin{equation}
\lambda = \sqrt{1+N \lambda_p^2} \mbox{\hspace{3ex}and\hspace{2ex}}
\lambda_\textrm{eff} =  \frac{N \lambda}{N+\lambda -1} \,.
\end{equation}
This is simply seen by scaling the dimensionally-scaled radius $\bar{r}$\,:
\begin{equation}
\bar{r}_{\rm eff} = \sqrt{\lambda_{\rm eff}} \, \bar{r}
\end{equation}
and scaling the wave function by the multiplier $1/\sqrt{\lambda_{\rm eff}}$ from the change of variables
\begin{equation}
d\bar{r} = \frac{1}{\sqrt{\lambda_{\rm eff}}} \, d\bar{r}_{\rm eff}
\end{equation}
which gives the equation for the universal curve of the density profile as
\begin{equation} \label{eq:Nbarreff}
\mathcal{N}(\bar{r}_{\rm eff}) = \frac{2 \, D^{\frac{D}{2}} } {\Gamma \hspace{-0.4ex} \left( \frac{D}{2} \right) }
\,\, \bar{r}_{\rm eff}^{D-1} \, \exp{ \left(  - D \,\,\bar{r}^2_{\rm eff} \right)   } \,,
\end{equation}
where
\begin{equation}
\int_0^\infty \mathcal{N}(\bar{r}_{\rm eff}) d\bar{r}_{\rm eff} = 1\,.
\end{equation}

As $D \rightarrow \infty$\,, the number density of Eq.~(\ref{eq:Nbarr}) becomes more and more strongly peaked at
$(\bar{r}_{\rm eff})_\infty$ determined from
\begin{equation}
\left. \frac{d \mathcal{N}(\bar{r}_{\rm eff}) }{ d\bar{r}_{\rm eff} } \right|_{D \rightarrow \infty} = 0\,,
\end{equation}
and thus the peak of the number density occurs precisely at the large-$D$ radius parameter
\begin{equation} \label{eq:rbareffinfinity}
(\bar{r}_{\rm eff})_\infty = \frac{1}{\sqrt{ 2 } } \,.
\end{equation}
%
\subsubsection{Series expansion}
As in Ref.~\onlinecite{GFpaper}, we introduce the dimensionally-scaled displacement coordinate
\begin{equation}
\bar{r}_{\rm eff} = (\bm{r}_{\rm eff})_\infty + \delta^{\frac{1}{2}} \bar{r}'_{\rm eff} = \frac{1}{D^{\frac{1}{2}}} ( \bar{r}'_{\rm eff} + D^{\frac{1}{2}} (\bar{r}_{\rm eff})_\infty )
\end{equation}
so that
\begin{equation}
d \bar{r}_{\rm eff} = \frac{1}{D^{\frac{1}{2}}} d \bar{r}'_{\rm eff}\,.
\end{equation}
Thus
\begin{equation} \label{eq:Nbarrexp}
\mathcal{N}(\bar{r}'_{\rm eff}) = \frac{2 } {\Gamma \hspace{-0.4ex} \left( \frac{D}{2} \right) }
\,\, ( \bar{r}'_{\rm eff} + D^{\frac{1}{2}} (\bar{r}_{\rm eff})_\infty )^{D-1} \, \exp{ \left(  -  \, ( \bar{r}'_{\rm eff} + D^{\frac{1}{2}} (\bar{r}_{\rm eff})_\infty )^2 \right)   } \,,
\end{equation}
where
\begin{equation} \label{eq:Nexact}
\int_{- \sqrt{D} \, (\bar{r}_{\rm eff})_\infty}^\infty \mathcal{N}(\bar{r}'_{\rm eff}) \, d\bar{r}'_{\rm eff} = 1\,.
\end{equation}

To derive the dimensional expansion of Eq.~(\ref{eq:Nbarrexp}), let's first consider expanding
$(\bar{r}'_{\rm eff} + D^{\frac{1}{2}} (\bar{r}_{\rm eff})_\infty )^{D-1}$\,. We derive
\renewcommand{\jot}{0.5em}
\begin{eqnarray}
(\bar{r}'_{\rm eff} + D^{\frac{1}{2}} (\bar{r}_{\rm eff})_\infty )^{D-1}
& = & \left( \frac{D}{2} \right)^{\frac{D-1}{2}}  \, \left(  1 + \delta^{\frac{1}{2}} 
\left(  \frac{\bar{r}_{\rm eff}^{\prime \, 3} }{ 3 (\bar{r}_{\rm eff})_\infty^3}  -  \frac{\bar{r}'_{\rm eff} }{ (\bar{r}_{\rm eff})_\infty }    \right)  + O(\delta)      \right) 
\nonumber \\
&& \times \, \exp{\left( \frac{D^{\frac{1}{2}} \, \bar{r}'_{\rm eff} }{ (\bar{r}_{\rm eff})_\infty}   \right) }
\, \exp{\left(  - \, \bar{r}_{\rm eff}^{\prime \, 2}  \right)  } \,. \label{eq:rbpdm1}
\end{eqnarray}
\renewcommand{\jot}{0em}
Likewise we also have
\begin{equation} \label{eq:explrbp}
\exp{ \left(  - \, ( \bar{r}'_{\rm eff} + D^{\frac{1}{2}} (\bar{r}_{\rm eff})_\infty )^2 \right)   }
= \exp{ \left(   - \frac{D}{2}  \right) } \, \exp{ \left(  - D^{\frac{1}{2}} \frac{\bar{r}'_{\rm eff}}{(\bar{r}_{\rm eff})_\infty} \right) }
  \, \exp{  \left(   -  \, \bar{r}_{\rm eff}^{\prime \, 2} \right) }\,.
\end{equation}
Equations~(\ref{eq:rbareffinfinity}), (\ref{eq:rbpdm1}) and (\ref{eq:explrbp}), along with
%
\begin{equation} \label{eq:sqrtoneOGamma}
\sqrt{\frac{1}{\Gamma\left(\frac{D}{2}\right)} } = \frac{2^{\frac{D-2}{4}}\exp{(\frac{D}{4}}) }{\sqrt[4]{\pi} D^{\frac{D-1}{4}} }
+ O(\delta) \,,
\end{equation}
give us the
result we are after, namely
\begin{equation} \label{eq:Ndensharm1}
\mathcal{N}(\bar{r}'_{\rm eff}) = \left(  1 + \delta^{\frac{1}{2}} \, \sqrt{2} \, \left(  \frac{2 \, \bar{r}_{\rm eff}^{\prime 3} }{3 } 
- \bar{r}'_{\rm eff}  \right)  +  O(\delta)  \right)
\left( \frac{2 } {\pi} \right)^{\frac{1}{2}} \, \exp{ \left(  -2 \, \bar{r}_{\rm eff}^{\prime 2}   \right)  } \,,
\end{equation}
where through order $\delta^{\frac{1}{2}}$ the normalization condition
\begin{equation}
\int_{-\infty}^\infty \mathcal{N}(\bar{r}'_{\rm eff}) \, d\bar{r}'_{\rm eff} = 1
\end{equation}
is still satisfied.

As we noted after Eq.~(\ref{eq:normbar}), the density profile for
harmonically-interacting particles in a harmonic confining potential follows a universal curve for any
$N$ or interparticle interaction strength $\lambda_p = \omega_p / \omega_t$\,. Although the density profile has this property,
the same cannot be said for the wave function for $N$\,, harmonically-interacting particles in a harmonic
confining potential. For example, many terms in the wave function through first order in $\delta^{\frac{1}{2}}$
(see Eqs.~(24), (25) and (26) of Ref.~\onlinecite{wavefunction1}) are zero for the free trap ($\lambda_p=0 \rightarrow \lambda=1 \rightarrow \lambda_{\rm eff} = 1$\,,
$\bar{r}_\infty = 1/ \sqrt{2}$\,, and $\gamma_\infty = 0$). Thus there is no simple scaling between the wave function for non-interacting particles in a harmonic confining potential and the wave function for harmonically-interacting particles in a harmonic confining potential.


\begin{thebibliography}{99}
\bibitem{multigraph}
Strictly speaking, this is a ``loop multigraph''. The definition of a graph
  does not allow for multiple edges between a pair of vertices nor a ``loop''
  edge with common endpoints.
\end{thebibliography}

%
\end{document}